\documentclass[prc,preprint,nofootinbib,preprintnumbers,
superscriptaddress,tightenlines
]{revtex4-1}

\usepackage{amsmath,mathrsfs}    
\usepackage{graphicx}   
\usepackage{subfigure}
\usepackage{color}
\usepackage[colorlinks=true,linkcolor=blue,citecolor=blue, urlcolor=blue]{hyperref}   
\usepackage{bm} 
\usepackage{ulem}
\usepackage[inline]{enumitem}
\usepackage{soul}
\usepackage{graphicx}
\usepackage{amsmath,amssymb}
\usepackage{soul}
\usepackage{textcomp}
\usepackage{hyperref}
\usepackage{subfigure}
\usepackage[toc,page]{appendix}

\def\half{{\textstyle\frac{1}{2}}}

\def\p{{\bm p}}
\def\x{{\bm x}}

\def\q{{\bm q}}
\def\k{{\bm k}}

\def\q{{\bm q}}

\def\bx{{\overline{x}}}

\def\k{{\bm k}}
\def\F{{\mathcal H}}

\def\st{\begin{equation}}
\def\stp{\end{equation}}
\def\bg{\begin{eqnarray}}
\def\nd{\end{eqnarray}}
\def\Eq#1{eq.~(\ref{#1})}

\def\eq#1{(\ref{#1})}
\def\app#1{Appendix~\ref{#1}}
\def\Fig#1{Fig.~\ref{#1}}

\def\Sect#1{Sect.~\ref{#1}}

\def\llangle{\left\langle}
\def\rrangle{\right\rangle}
\def \bes {\begin{subequations}}
\def \ees {\end{subequations}}

\def \chih {{\chi_A}}
\def \chione {{\chi_1}}

\def \one{\mathbb I}
\def \omq{\omega_q}
\def\tr{\mathrm{tr}}
\def\Obulk{ {\mathcal O}_{\rm bulk}}
\def\Gammachem{\Gamma_{\rm chem}}
\def\diss{diss}
\def\MM{\mathcal M}
\def\chemconst{{\chi_0}}
\def\ie{i.e.~}
\def\al{{\mathscr L}}
\def\ar{{\mathscr R}}
\def\tr{{\mathrm{tr}}}

\newcommand{\LL}{{\scriptscriptstyle L}}
\newcommand{\RR}{{\scriptscriptstyle R}}
\newcommand{\VV}{{\scriptscriptstyle V}}
\newcommand{\Aa}{{\scriptscriptstyle A}}
\begin{document}


\title{Transport and hydrodynamics in the chiral limit}

\author{Eduardo Grossi}
\email[]{eduardo.grossi@stonybrook.edu}
\affiliation{Department of Physics and Astronomy, Stony Brook University, Stony Brook, New York 11794, USA}
\author{Alexander Soloviev}
\email[]{alexander.soloviev@stonybrook.edu}
\affiliation{Department of Physics and Astronomy, Stony Brook University, Stony Brook, New York 11794, USA}
\author{Derek Teaney}
\email[]{derek.teaney@stonybrook.edu}
\affiliation{Department of Physics and Astronomy, Stony Brook University, Stony Brook, New York 11794, USA}
\author{Fanglida Yan}
\email[]{yan.fanglida@stonybrook.edu}
\affiliation{Department of Physics and Astronomy, Stony Brook University, Stony Brook, New York 11794, USA}

\preprint{MIT-CTP/5042}
\date{\today}

   \begin{abstract}
      We analyze the evolution of hydrodynamic fluctuations for QCD matter below $T_c$ in the chiral limit, where the pions (the Goldstone modes) must be treated as additional non-abelian superfluid degrees of freedom, reflecting the broken $SU_L(2) \times SU_R(2)$ symmetry of the theory.  
   In the presence of a finite pion mass $m_{\pi}$, the hydrodynamic theory is ordinary hydrodynamics at long distances, and superfluid-like 
   at short distances. The presence of the superfluid degrees
   of freedom then gives specific contributions to the bulk viscosity, the 
   shear viscosity, and diffusion coefficients of the ordinary theory 
   at long distances which we compute. This determines,  in some cases, the leading dependence of the transport parameters of QCD on the pion mass. We analyze the predictions of this computation, as the system approaches the $O(4)$ critical point. 
   \end{abstract}


\maketitle

\clearpage

\section{Introduction}
\label{intro}

Viscous hydrodynamics, based on the conservation of energy and momentum,  
is remarkably successful at describing a wide range of correlations
observed in heavy ion collisions and  has become a kind of ``standard model'' for 
heavy ion events~\cite{Teaney:2009qa,Romatschke:2017ejr}.  Hydrodynamics is a long wavelength effective theory 
which captures the underlying symmetries of the microscopic theory.  In QCD this symmetry 
is approximately $U(1) \times SU_{L}(2) \times SU_{R}(2)$, which below a transition temperature is 
broken  to  $U(1) \times SU_V(2)$ when the chiral condensate $\llangle \bar q q\rrangle$ develops. 
In the chiral  limit $m_q \rightarrow 0$ this symmetry is exact and is associated
with strictly massless Goldstone modes. In the chiral limit, and below the transition temperature, these modes should be added
to the usual hydrodynamic modes associated with energy momentum and charge conservation, leading
to an effective theory which is  analogous to a non-abelian superfluid~\cite{Son:1999pa,Son:2002ci}.

\begin{figure}
 \includegraphics[width=0.8\textwidth]{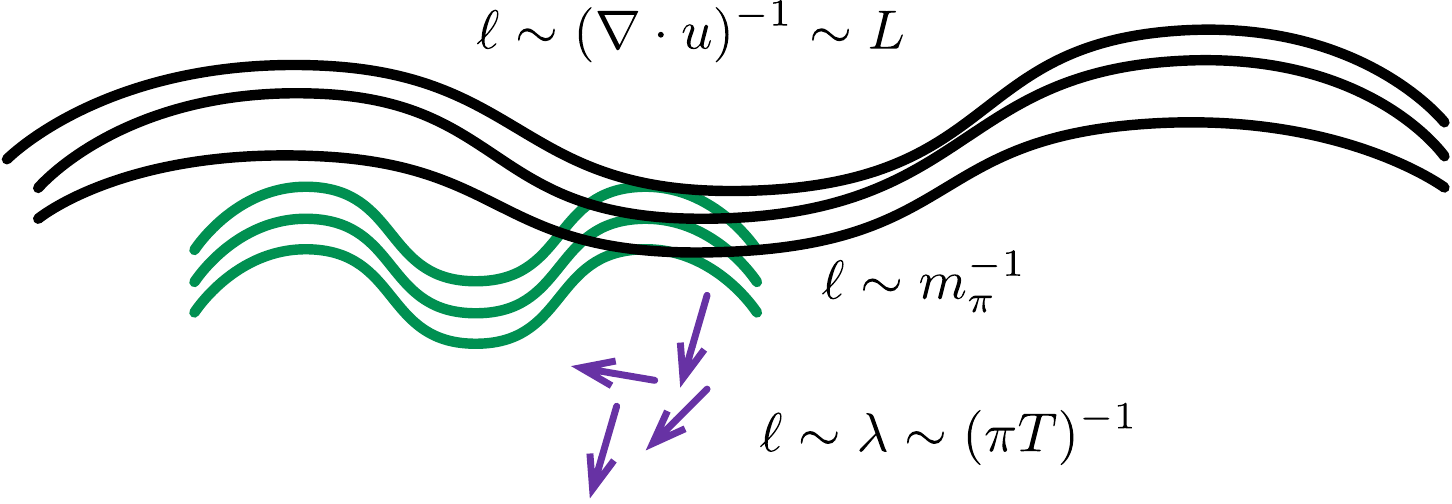}
 \caption{\label{cartoon} 
    Long wavelength modes (black lines) with $\ell \sim  (\nabla\cdot u)^{-1} \sim L$
    are described with ordinary hydrodynamics.
To model wavelengths  of order $\ell \sim (m_\pi)^{-1}$, the effective theory must treat  the soft pion modes explicitly (green lines). 
These modes can be described with a non-abelian superfluid theory, due to the fact that the pions are Goldstone bosons. Finally, the microscopic degrees of freedom (purple arrows), which include typical pions with $p \sim  \lambda^{-1} \sim \pi T$ and other hadronic states,  determine the thermodynamic and dissipative parameters of the superfluid.  The superfluid modes leave calculable imprints on the transport parameters of the ordinary fluid.
 }
\end{figure}

In the presence of a finite quark mass, chiral symmetry is no longer an exact symmetry, and at long
distances the appropriate effective theory is ordinary hydrodynamics. 
Nevertheless, 
the quark mass is small and one can reasonably ask whether the superfluid
effective theory leaves any imprint on the evolution of the system.  At finite quark mass the theory 
should be superfluid-like for modes with wavelength $\ell \sim m_{\pi}^{-1}$  and should asymptote
to ordinary hydrodynamics for $\ell \gg m_{\pi}^{-1}$, with the superfluid modes correcting the ordinary transport
coefficients of QCD.  These corrections are determined by the dissipative parameters of the superfluid theory.
One of our goals in this paper is to present these corrections, which (in some regimes) are the leading contributions of the pion mass to the transport coefficients of QCD.    
The physical picture is summarized in \Fig{cartoon}.

This is a particularly current time to consider chiral physics. Work from the lattice~\cite{Ding:2019prx,Borsanyi:2020fev}
provides evidence that finite temperature QCD in the real world approximately exhibits the scaling behavior of $O(4)$ symmetric models. It is thus natural to think that passing close to the chiral phase transition, the phase of the condensate will get generated as the condensate builds. 
The pions emitted in this way will have  small momenta and therefore can  escape the system unscathed, possibly leaving a soft pion signal of the chiral dynamics in the detector.

Observation of soft pions has been difficult.  Fortunately, an upgrade is
underway to the ITS detector at ALICE \cite{ALICE} that could provide a wider
window into low $p_T$ particles, especially pions. This can shed light on the
physics driven by the chiral phase transition. There are many interesting
scenarios to explore using soft pions, such as the Bose-Einstein condensation
of pions \cite{PhysRevC.91.054909}, or Disoriented Chiral Condensate (DCC)
\cite{Blaizot:1992at,Rajagopal:1993ah,Mohanty:2005mv}. The standard observable
proposed to detect the soft dynamics of pions induced by the chiral phase
transition is the multiplicity ratio of charged pions with the neutral one
\cite{Rajagopal:1993ah}. Another possible source of information about the
chiral phase transition can be expected to manifest itself in the correlation
functions between charged pions \cite{Adam:2015pbc}.


Previous work includes a model where a fluid coupled to the chiral condensate (and the gluon condensate) was considered \cite{Herold:2014vba,Nahrgang:2011mv,Nahrgang:2011mg,Nahrgang:2018afz} in the context of computing multiplicity fluctuations near the phase transition. In the chiral sector, the model only captured the evolution of the order parameter, neglecting the dynamics of the $SU(2)$ phase. In the present model, we explicitly consider the dynamics of the phase in the broken phase.

The basic non-Abelian superfluid equations of motion were written down by D.~Son many years ago~\cite{Son:1999pa}. These
equations were extended to include dissipation and at a linearized level the effect of 
a finite quark mass~\cite{Son:2002ci,Pujol:2002na}. Formal developments by Jain (building on~\cite{Bhattacharyya:2010yg,Bhattacharyya:2012xi,Haehl:2014zda})
have considerably clarified the general structure of the equations of motion~\cite{Jain:2016rlz}.  After reviewing the equations in \Sect{eom},  
we will describe the behavior of the hydrodynamic correlation functions in \Sect{kubo}, which can be used
to determine how the transport coefficients of QCD depend on the pion mass.  Finally in \Sect{discussion}, we discuss the expected scaling behavior of the computed transport coefficients in the vicinity of the critical point.  

\section{The hydrodynamic equations close to the chiral limit}
\label{eom}

This section briefly reviews the equations of motion discussed in \cite{Son:1999pa,Son:2002ci,Pujol:2002na,Floerchinger:2016gtl,Jain:2016rlz}.
Chiral symmetry breaking and its associated effective Lagrangian are reviewed with precision and clarity in~\cite{Scherer:2002tk,donoghue1994dynamics}.

\subsection{Ideal  hydrodynamics} 

The hydrodynamic theory is based on the conserved charges and phases associated with broken  $SU_L(2) \times SU_R(2)$. 
The invariance of the theory yields two conserved currents with independent left and right isospin rotations
\begin{align}
   J_L^{\mu} =& (J_{L})_{a}^{\mu} t^a \,,  \\
   J_R^{\mu} =& (J_{R})_{a}^{\mu} t^a \, ,
\end{align}
where the generators 
$t^a$ are proportional to the  Pauli matrices, with trace normalization $\tr[t^a t^b] = T_F \delta^{ab}$.  The equilibrium state is characterized
by the chiral condensate $\Sigma \equiv \llangle \bar q_R q_L \rrangle$, which transforms 
as $\Sigma \rightarrow g_{L} \Sigma g_{R}^\dagger$ under a chiral rotation.
At each point in space and time, the local value chiral condensate $\Sigma \equiv \llangle \bar q_R q_L \rrangle(x) $ is rotated
relative to a reference state $\Sigma^{(0)} = \sigma(T) \one$  
by an axial rotation, where   $g_L=g_R^\dagger = \xi$. 
The phase  $\xi=\exp(i\varphi)$ is parametrized by the pion field\footnote{
   In chiral perturbation theory $\varphi=\pi/F$ where at leading order $F \simeq f_{\pi}$.
}
$\varphi = \varphi_a(x) t^{a}$.
Since under independent left and right rotations $\Sigma  \rightarrow  g_{L} \Sigma g_R^{\dagger}$,
the condensate may be written $\Sigma = \sigma U$, where $U \equiv \xi^2$ is the phase, \ie the unitary matrix that is traditionally
used to parametrize the chiral Lagrangian. For the purposes of this paper, we will take $\sigma$ to be a 
constant. Fluctuations of $\sigma$ will be considered in future works.

As just discussed,  the system at a point $x$ is rotated relative to a reference state by an $SU_L(2) \times SU_R(2)$ rotation  parametrized by $(U_L(x), U_R(x))$.
The left and right chemical potentials are related to  time derivatives of these rotation matrices~\cite{*[{This definition and choice of variables for the fluid action was inspired by }][] Glorioso:2018wxw},
\begin{align}
   \mu_{L} & \equiv i u^{\mu} D_{\mu}^{\al} U_{L}  U^{\dagger}_{L}  = i u^{\mu} \partial_\mu U_{L} U_{L}^\dagger + u^{\mu} \al_{\mu}, \\
   \mu_{R} & \equiv i u^{\mu} D_{\mu}^{\ar} U_{R}  U^{\dagger}_{R}  = i u^{\mu} \partial_\mu U_{R} U_{R}^\dagger  + u^{\mu} \ar_{\mu},  
\end{align}
where $\al_\mu$ and $\ar_\mu$ are external left and right gauge fields, and 
the flow velocity $u^{\mu}$ is timelike normalized, $u^{\mu}u_{\mu}=-1$.
These relations are called Josephson constraints in the superfluid theory.
The chiral condensate $\Sigma$ is given by $\Sigma = U_{L} \, (\sigma \mathbb{I})\,  U_R^{\dagger}$, 
and thus the unitary matrix $U$  is simply  $U = U_{L} U_R^\dagger$.  In constructing 
the chiral Lagrangian  it is customary to introduce the left and right handed currents, $L_{\mu}\equiv i UD_{\mu} U^\dagger$
and $R_{\mu}\equiv i U^\dagger D_{\mu} U=-U^\dagger L_{\mu} U$, defined with the appropriate covariant derivatives 
\begin{align}
 D_{\mu} U&= \partial_{\mu}U-  i\al_{\mu}U+iU\ar_{\mu}, \\
 D_{\mu} U^{\dagger}&= \partial_{\mu}U^{\dagger}+  iU^{\dagger}\al_{\mu}-i\ar_{\mu}U^{\dagger}.
\end{align}
Using these definitions,
we  find that the zeroth component of the left and right handed currents are related to the difference in chemical potentials
\begin{align}
   -u^{\mu} L_{\mu} =& i u^{\mu} D_{\mu} U U^{\dagger}  = \mu_{L} - U \mu_R  U^{\dagger},  \\
   -u^{\mu} R_{\mu}   =& i u^{\mu} D_{\mu} U^\dagger U  = \mu_{R} - U^\dagger \mu_L  U. 
\end{align}
These are analogous to the $U(1)$ superfluid Josephson relation, where $-u^{\mu} \partial_{\mu}\varphi = \mu$. 

To quadratic order in $\mu_L$, $\mu_R$, and $L_{\mu}$,  the $SU_L(2){\times}SU_R(2)$ 
invariants are $\mu_L^2 + \mu_R^2$, $(\mu_L - U \mu_R U^{\dagger})^2$ and $L_{\mu} L^{\mu}$.
Thus, a general action for ideal hydrodynamics  close to  the chiral limit is 
\begin{align}
  S 
  =&\int \text{d}^4 x   \sqrt{-g} (p(T)+\mathcal{L}_{\text{superfluid}}),
\end{align}
where\footnote{Our normalization constants here are chosen so that the vector chemical potential is 
an average of the left and right chemical potentials, while the vector current is  a sum of the left and right
 currents, so that $\mu_L \cdot J_{L} + \mu_R \cdot J_R = \mu_V \cdot J_V + \mu_A \cdot J_A$. Thus, the $O(4)$ symmetric term reads
 $\tfrac{1}{4} \chemconst(\mu_L^2 + \mu_R^2) =  \tfrac{1}{2} \chemconst (\mu_V^2 + \mu_A^2)$.}
\begin{multline}
   \label{superfluidlagrange}
   \mathcal{L}_{\text{superfluid}}= \frac{1}{ 4 T_F} 
   \tr \big[\chemconst   \, (\mu_L^2+ \mu_R^2) \big]  + \\
    \frac{1}{8 T_F} \tr \big{[} \chione \, u^{\mu}u^{\nu} D_\mu U  D_\nu U^{\dagger}   -  f^2  D_\mu  U D^\mu U^\dagger 
+  f^2 m^2 ( U \mathcal{M}^\dagger  +  \mathcal{M} U^\dagger) \big{]},
\end{multline}
%
%
%
%
Here we are tracing over isospin. $p(T)$ is the pressure as a function of temperature,
 defined through the vector $\beta^{\mu}$,  i.e.
$ T{\equiv}(-\beta^\mu g_{\mu\nu} \beta^\nu)^{-1/2}$   
with $\beta^\mu=\frac{1}{T}u^\mu$.
$\mathcal{M}$ is a fixed matrix which can be taken to be unity, and is responsible for the explicit breaking of chiral symmetry. Note that $m$ refers to the screening mass, which is directly 
related to the pole mass\footnote{In the finite temperature chiral perturbation theory 
   literature the susceptibility of the superfluid component, $f^2$, is called the spatial pion decay constant, $f_s^2$~\cite{Toublan:1997rr,Schenk:1993ru}.
   The total axial charge susceptibility, $\chi_A \equiv \chi_0 + \chi_1 + f^2$, 
   is called temporal pion decay constant,  $f_t^2$. The pion velocity $v^2\equiv f^2/\chi_A$.
   In the AdS superfluid literature, the susceptibility 
of the normal component, $\chi_A^{\rm nrm} \equiv \chi_0 + \chi_1$,  is called $\chi$~\cite{Bhattacharyya:2010yg,Jensen:2012jh}. },
  $m_p^2 \equiv v^2 m^2$~\cite{Son:2002ci}. 
The coefficients $\chi_0, \chi_1, f$ and $m$ are functions of the temperature.
For the purposes of this paper, ultimately we will work around a Minkowski background, $g_{\mu\nu}=\eta_{\mu\nu}$, and also turn the gauge fields off, $\al_\mu=\ar_\mu=0$. 
Similar Lagrangians considering $U(1)$ superfluids, and $U(1)$ vector and axial currents coupled to gauge fields, including a discussion about anomalies, can be found in \cite{Jensen:2012jh,Jensen:2013vta}.

The hydrodynamic equations are given by the conservation of the energy momentum tensor~\cite{Son:1999pa},
\begin{align}
\nabla_\mu T^{\mu\nu}&=0,
\end{align}
with
\begin{align}
   \label{Tmunudef}
    T^{\mu\nu}=\frac{2}{\sqrt{-g}}\frac{\partial \mathcal{L}\sqrt{-g}}{\partial g_{\mu\nu}} 
&=\varepsilon_U u^\mu u^{\nu}
    + \Delta^{\mu\nu} p_{U}
    + \frac{f^2}{8 T_F}\tr \left(  D^\mu  U D^\nu U^\dagger +  D^\nu  U D^\mu U^\dagger\right),
\end{align}
where $\Delta^{\mu\nu} \equiv \eta^{\mu\nu} + u^{\mu} u^{\nu}$ is the projector onto the local rest frame, the redefined pressure is
\begin{align}
p_U &= p(T)+\mathcal{L}_{\text{superfluid}},
\end{align}
and the redefined energy density is given by a  Legendre transform of $p_U$
\begin{align}
   \label{epsilonU}
   \varepsilon_U=\varepsilon(T)+ \left( -1 +T\frac{\partial}{\partial T} + \mu_{L}^a \frac{\partial}{\partial \mu_L^a} + \mu_R^a \frac{\partial}{\partial \mu_R^a }  \right)\mathcal{L}_{\text{superfluid}}  .
\end{align}
When determining $\varepsilon_U$, the Lagrangian should be considered a function of 
the independent variables $T$, $\mu_L$, $\mu_R$, $\tr(\partial_{\mu} U \partial^{\mu} U^\dagger)$, and 
${\mathcal M}^\dagger U + {\mathcal M}  U^\dagger$, and thus the $\chi_1$ term in the $\mathcal L$ should be written:
\st
\chi_1 \tr[ u^{\mu} u^{\nu} D_\mu U D_{\mu} U^{\dagger}]  = \chi_1 \tr[ (\mu_L - U \mu_R U^{\dagger})^2 ] \, .
\stp

The ideal equations of motion of the chiral degrees of freedom read 
\begin{align} 
   D_\mu^{\al} J_{L}^\mu  &= -\frac{f^2 m^2  }{8} i (U \mathcal{M}^\dagger - \mathcal{M} U^\dagger),  \\
   D_\mu^{\ar} J_{R}^\mu &= +\frac{f^2 m^2  }{8} i (\mathcal{M}^\dagger U -  U^\dagger \mathcal{M}),
\end{align}
where the left and right currents are given by
\begin{align}
   J_{L}^\mu&=\frac{\delta S}{\delta \al_{a\mu}} t^{a}  = \frac{1}{2} \chemconst \mu_{\LL} u^{\mu}
   + \frac{1}{4} \chi_1 (\mu_\LL - U \mu_\RR U^{\dagger}) u^{\mu}    +  \frac{1}{4} f^2 L^{\mu} \, ,  \\
 J_{R}^\mu&=\frac{\delta S}{\delta \ar_{a \mu}} t^a  = \frac{1}{2} \chemconst \mu_{\RR}  u^{\mu}
 +\frac{1}{4}  \chi_1 (\mu_{\RR} - U^\dagger \mu_\LL U) u^{\mu}   +   \frac{1}{4} f^2 R^{\mu} \, .
\end{align}

Note that the conserved isovector current for $\mathcal M$ unity (i.e. the real world) is given by $\tilde J_V^\mu=J^\mu_L+\mathcal M J^\mu_R \mathcal M^\dagger$.  It is also useful
to consider  $J_{V}^{\mu} \equiv J^{\mu}_L + U J^{\mu}_R U^\dagger$, the associated chemical potential 
$ \mu_V \equiv (\mu_L + U \mu_R U^\dagger)/2$,
and corresponding axial definitions,  $J_A^{\mu} \equiv J^{\mu}_{L}
- U J_{R}^{\mu} U^\dagger$ and $ \mu_A \equiv (\mu_\LL - U \mu_\RR U^\dagger)/2$,
which can be interpreted as the projection of the current and chemical potentials  
onto the iso-vector and iso-axial-vector directions as seen from the reference state~\cite{Jain:2016rlz}. These projected currents read 
\begin{align}
   J_\VV^\mu  =& \chi_0 \,   \mu_\VV \,  u^{\mu}, \\
   J_{\Aa}^{\mu}  =& \chi_A^{\rm nrm} \,  \mu_\Aa  \, u^{\mu}  + \half f^2 L^{\mu} \, .
\end{align}
where we have defined $\chi_A^{\rm nrm} \equiv \chi_0 + \chi_1$. 
The form of the iso-vector current leads us to
identify $\chi_0$ as the iso-vector susceptibility. The iso-axial-vector current
consists of a normal component with susceptibility $\chi_A^{\rm nrm}$ and a
superfluid component with susceptibility $f^2$. The total iso-vector-axial charge density,
$-u_{\mu}  J^{\mu}_{\Aa}$, is  the total axial susceptibility, 
$\chi_{\Aa} \equiv  (\chi_A^{\rm nrm} + f^2)$, times the axial chemical potential, $ \mu_{\Aa}$. 

\subsection{Viscous corrections, entropy production, and noise}

\subsubsection{Viscous corrections and entropy production}

The equations that we have considered so far are ideal. 
We will be interested in computing the viscous corrections to the energy momentum tensor due to the viscous effects of the chiral sector.  This section extends \cite{Pujol:2002na,Jain:2016rlz} by including the mass terms (which are very important in practice), and \cite{Son:2002ci} by treating the theory non-linearly, which leads to an additional constraint. 

To this end we write
\begin{align}
T^{\mu\nu}=&T^{\mu\nu}_{\text{ideal}} + \Pi^{\mu\nu} \, , \\
J^{\mu}_L=& J^{\mu}_{L,\text{ideal}} + q_{L}^{\mu}  \, , \\
J^{\mu}_R=& J^{\mu}_{R,\text{ideal}} + q_{R}^{\mu}  \, . 
\end{align} 
and allow for a viscous correction to the Josephson constraint
\st \label{josephson}
 -\tfrac{1}{2} u^{\mu} L_{\mu}  = \mu_A  +  \mu_{A}^{\diss} \, .
\stp
The phenomenological currents should be proportional to the 
strains, and entropy production should be positive.  We may
choose the Landau frame such that 
\st
   u_{\mu} q_{L}^{\mu} = u_{\mu} q_R^{\mu} = u_{\mu} \Pi^{\mu\nu} = 0 \,.
\stp
 Finally, we will 
 further decompose the stress tensor into shear and bulk strains
\st
       \Pi^{\mu\nu} = \pi^{\mu \nu} +  \Pi \Delta^{\mu\nu} \, ,
\stp
where $\pi^{\mu}_{\;\mu} =0$.

The entropy is defined by the energy density $e_U$, pressure $ p_U$ and left and right density $n_{L/R} $:
\st \label{entropy}
s_U =  \frac{ e_U + p_U - \mu_L \cdot n_L - \mu_R \cdot n_R } {T},
\stp
where we introduced the shorthand, $\mu_L \cdot n_L = {\rm tr}[ \mu_L n_L] /T_F$.
A straightforward analysis of entropy production using
the equations of motion as seen in Appendix \ref{app:ent} yields
\st
\label{eq:entropyproduction}
 \partial_{\mu} \left( s_U u^{\mu} -  \mu_L \cdot q_{L}^{\mu} 
 - \mu_R \cdot q_R^{\mu} \right) 
 =  -\Pi^{\mu\nu} \, \partial_{\mu} \beta_{\nu}  
  -    q_L^{\mu} \cdot \partial_{\mu} \hat\mu_L  - q_R^{\mu} \cdot \partial_{\mu} \hat\mu_{R} 
   -\frac{\mu_{A}^{\diss}}{T} \cdot  \Theta_s  \, ,
\stp
where 
$\hat\mu \equiv \mu/T$.
The superfluid expansion scalar in this expression is 
\st
\label{eq:thetasdef}
\Theta_s = \left[ \partial_{\mu} \left( \frac{f^2}{2} L^{\mu} \right)  + \frac{ f^2 m^2}{4} \,i( U \mathcal M^\dagger - \mathcal M U^\dagger) \right]  \, ,
\stp
and is given by the variation of the ideal action leaving the temperature,
$\mu_V$, and $\mu_A$ fixed 
\st
\label{eq:thetasviaS}
\left(\delta S\right)_{\beta,\mu_V, \mu_A} \equiv \int d^4x \,\left(-\tfrac{i}{2} \delta U U^\dagger \right) \cdot \Theta_s  \, .
\stp

Requiring positivity of entropy production leads in
the tensor sector to 
\begin{align}
    \pi^{\mu\nu} =& -\eta^{(0)}  \, \sigma^{\mu\nu}  \quad \mbox{with} \quad   \eta \geq 0 \, .
\end{align}
In the scalar sector there are two structures, leading to 
the constitutive relations
\begin{align}
   -\Pi &=  \zeta^{(0)}  \,  \nabla \cdot u   + \zeta^{(1)} \, \mu_A \cdot \Theta_s  \, ,  \\ 
 -\mu_A^{\rm diss}  &=   \zeta^{(1)} \, \mu_A \, \nabla \cdot u + \; \zeta^{(2)} \phantom{\cdot} \Theta_s \, .    
\end{align}
For the quadratic form, $-(\Pi \, \nabla \cdot u +  \mu^{\diss}_A \cdot \Theta_s) $, to be non-negative we must have
\st
    \zeta^{(0) } \geq 0, \qquad \zeta^{(2) } \geq 0, \qquad \zeta^{(0)} \zeta^{(2)} - (\zeta^{(1)})^2 \mu_{A}^2  \geq  0 \, .
\stp
In the vector-sector we have\footnote{ $q_V^{\mu} $ is not strictly speaking a vector. Under parity 
it is transformed to   $q_{V}^{\mu} \rightarrow U^{\dagger} q_{V}^{\mu} U$.  
The quantity $\xi^\dagger q_{V}^{\mu} \xi$ is a vector in a strict sense. The terms in \Eq{vectoreqn} are grouped according to familiar covariant derivatives
of chiral perturbation theory. 
In particular,
rotating $\mu_V$ to the refernce state $\mu_{V}^\xi \equiv \xi^{\dagger} \mu_V \xi$,  and defining the vector field, 
$v_{\mu}\equiv-i \left(\xi^\dagger \partial_{\mu}\xi + \xi\partial_{\mu}\xi^\dagger \right)$,
the covariant 
derivative is 
\st
d_\mu \mu_V^\xi \equiv \left(\partial_{\mu} + i v_{\mu} \right) \mu_{V}^{\xi} = \xi^{\dagger} \left( \partial_{\mu}\mu_V  - \tfrac{i}{2} \left[ L_{\mu}, \mu_V \right] \right) \xi \, .
\stp
Both $\mu_V^{\xi}$ and $d_{\mu} \mu_V^{\xi}$ are
directed in the unbroken iso-vector subgroup in the reference state, while $\partial_{\mu} \mu_V^{\xi}$ is not~\cite{Jain:2016rlz,donoghue1994dynamics}. 
}
\begin{subequations}
\begin{align}
q_V^{\alpha} =& -T\sigma_I^{(0)} \Delta^{\alpha\beta}(\partial_{\alpha} \hat \mu_L + U  \,\partial_{\beta} \hat \mu_R\,  U^\dagger )/2\,  , \\
\label{vectoreqn}
     =&   -T\sigma_{I}^{(0)} \,
   \Delta^{\alpha\beta} \left( (\partial_{\beta}\hat \mu_V - \tfrac{i}{2}  [L_{\beta}, \hat \mu_V])  + \tfrac{i}{2} [L_{\beta}, \hat \mu_A]  \right)  \quad  \mbox{with} \quad   \sigma_{I}^{(0)} \geq 0 \, .
   \end{align}
\end{subequations}
   In the pseudo-vector sector we have
   \begin{subequations}
   \begin{align}
q_A^{\alpha} =& -T\sigma_A \Delta^{\alpha\beta}(\partial_{\alpha} \hat \mu_L - U  \,\partial_{\beta} \hat \mu_R\,  U^\dagger)/2 \, , \\
     =&   -T\sigma_{A} \Delta^{\alpha\beta} \left( (\partial_{\beta}\hat\mu_A - \tfrac{i}{2} [ L_{\beta}, \hat \mu_{A} ]) + \tfrac{i}{2} [L_{\beta}, \hat \mu_V]  \right)   \quad \mbox{with} \quad  \sigma_{A} \geq 0 \, .
\end{align}
\end{subequations}

To summarize,  the superfluid theory contains the three transport coefficients
   of the normal theory, $\eta^{(0)}, \zeta^{(0)}, \sigma_I^{(0)}$.  In 
   addition, it contains two parameters, $\zeta^{(2)}$ and $\sigma_A$,
which describe the damping of the pions,
and which will be parameterized below by 
axial charge diffusion coefficient $D_A$ and the damping rate $D_{m}$.
Finally, the theory contains one additional coefficient, $\zeta^{(1)}$
that intrinsically couples the normal and pion sectors. 
This term involves two small parameters, the axial chemical potential $\mu_A$  and the viscous correction arising from $\nabla\cdot u$, and can probably be ignored in practice. 

It is notable that 
the two independent scalars  comprising the superfluid expansion
scalar,
$\partial_{\mu} (f^2 L^{\mu})$ and $U \mathcal M^\dagger -  \mathcal M U^{\dagger}$,  
must have the same dissipative coefficient, $\zeta^{(2)}$. This  
constraint, arising from entropy considerations, was not recognized in the linearized analysis of dissipation by Son and  Stephanov~\cite{Son:2002ci}, which leads to an additional transport coefficient in their theory\footnote{Specifically, we find
   that Son and Stephanov's coefficients $\kappa_1$ and $\kappa_2$  are both given by $\lambda_{m} = (\chi_A^{\rm nrm} v)^2  \zeta^{(2)}$.
}.  

This consequence of  entropy conservation simplifies the interpretation of 
the theory. 
For simplicity of presentation, we will set $\zeta^{(1)}$, $\sigma_A$, and $\sigma_I^{(0)}$ to zero, and call $\Gammachem\equiv 1/\zeta^{(2)}$.
(These constraints are all easily relaxed.)  
The chemical potential of the normal components is $\mu_A$, while
the chemical potential of the pion component is $\mu_A^{\varphi} \equiv -\tfrac{1}{2} u^{\mu} L_{\mu}$, and the two  potentials are trying to 
be made equal by the microscopic dynamics.
From this perspective it is not surprising  that the equations of motion 
can be easily rewritten in an intuitive form
   \label{chemicalrewrite}
\begin{align}
   \partial_{\mu} (n_L u^{\mu}) \,  - \, U\, \partial_{\mu} (n_R u^{\mu} ) \, U^{\dagger}
   =& - \Gammachem \left(\mu_A  - \mu_A^\varphi \right), \label{eq:normalpiece} \\
   \partial_{\mu} \left(\tfrac{f^2}{2} L^{\mu}  \right)  + \frac{f^2 m^2 }{4} i (U \MM^\dagger - \MM U^{\dagger} ) =& - \Gammachem \left(\mu_A^{\varphi} - \mu_A \right), \label{eq:pioneom} 
\end{align}
which clearly shows the chemical coupling between the pion equation of motion \eqref{eq:pioneom} and the normal axial components (i.e. hard pions and other hadronic states).

\subsubsection{Noise}
The analysis of entropy production also determines the thermodynamic noise in the system. Neglecting $\zeta^{(1)}$ 
for simplicity, the mean entropy production rate can be written
\st
\label{EntropyProductionAverage}
 \partial_{\mu} \left( s_U u^{\mu} -  \mu_V \cdot q_{V}^{\mu} 
 - \mu_A \cdot q_A^{\mu} \right) 
 =   \frac{ \pi^{\mu\nu} \pi_{\mu\nu} }{ 2 T \eta^{(0)}}  
  + \frac{\Pi^2}{ T \zeta^{(0) } }
  +   \frac{  q_V^{\mu} (q_{V})_\mu }{ T \sigma_V^{(0) }  }
  + \frac{ q_A^\mu (q_{A})_\mu}{ T \sigma_A} +  \frac{ (\mu_A^{\diss})^2 }{ T \zeta^{(2)} } \ .
\stp
In stochastic hydro,  noise should be added to each dissipative strain, i.e. 
\begin{subequations}
\begin{align}
   q_A^{\mu}  &\rightarrow  q_A^{\mu}  + \xi_{A}^{\mu} \, , \\
   \mu_A^{\diss}  &\rightarrow \mu_A^{\diss}  + \xi_{\mu_A}^{\diss}  \, ,
 \end{align}
 \end{subequations}
in addition to the familiar noises of ordinary hydrodynamics,  $\xi^{\mu\nu}_\pi$, $\xi_\Pi$, and $\xi_V^{\mu}$.
 The general theory of these fluctuations
 determines the variances of the noises from the equilibrium susceptibility matrix and the dissipative quadratic form for entropy production~\cite{Landau1,Landau2,Fox1,Fox2}.  
In the current case, these variances can be read off from
 the denominators of \eqref{EntropyProductionAverage}, i.e.
 \begin{subequations}
    \label{variancesfundamental}
 \begin{align}
    \llangle \xi^{\mu}_A(x) \xi_{A}^{\nu}(y) \rrangle &= 2 T \sigma_A \Delta^{\mu\nu} \delta(x -y)  \, , \\
    \llangle \xi_{\mu_A}^{\diss}(x) \xi_{\mu_A}^{\diss}(y) \rrangle &= 2 T \zeta^{(2)} \delta(x -y)  \, .
 \end{align}
 \end{subequations}
in addition to the usual variances for  $\xi^{\mu\nu}_\pi$, $\xi_\Pi$, and $\xi_V^{\mu}$.
In writing these formulae in this simple form it  
was important that we expanded $\mu^{\diss}_A$ in terms of the canonical conjugate of $U$,  as given by $\Theta_s$ in \eqref{eq:thetasviaS}. Otherwise, the form of the variances would also involve the equilibrium matrix of susceptibilities.

Finally, we note that in the presence of noise, the rearrangements of the equations of motion that lead to 
\eqref{eq:normalpiece} and \eqref{eq:pioneom} now give rise to a stochastic equation of chemical balance
\begin{align}
   \partial_{\mu} (n_L u^{\mu}) \,  - \, U\, \partial_{\mu} (n_R u^{\mu} ) \, U^{\dagger} 
   =& - \Gammachem \left(\mu_A  - \mu_A^\varphi \right)  -\xi_{\rm chem}, \label{eq:normalpiecewxi} \\
   \partial_{\mu} \left(\tfrac{f^2}{2} L^{\mu}  \right)  + \frac{f^2 m^2 }{4} i (U \MM^\dagger - \MM U^{\dagger} ) =& - \Gammachem \left(\mu_A^{\varphi} - \mu_A \right) + \xi_{\rm chem} , \label{eq:pioneomwxi}
\end{align}
where the chemical noise $\xi_{\rm chem}$ (which enters as $\Gammachem \, \xi_{\mu_A}^{\diss}$) satisfies 
the expected chemical fluctuation-dissipation relation
\begin{subequations}
\begin{align}
   \llangle \xi_{\rm chem}(x)  \xi_{\rm chem}(y) \rrangle &= (\Gammachem)^2 \llangle \xi_{\mu_A}^{\diss}(x) \, \xi_{\mu_A}^{\diss}(y) \rrangle \, , \\
   &=  2T\,\Gammachem \,\delta(x -y) \, ,
\end{align}
confirming the consistency of the interpretation.
\end{subequations}

\subsection{Linearized equations of motion}

Following Son and Stephanov~\cite{Son:2002ci}, we now parametrize the phase as $U=e^{2i\varphi}$ and linearize 
the equation of motion for $J_A^{\mu}$ together with the Josephson's constraint around global equilibrium: 
\begin{align}
   \partial_t(\chi_A^{\rm nrm} \mu_A)   + \nabla \cdot( -\sigma_A \nabla \mu_A  +  \vec{\xi}_A) =& - \Gamma_{\rm chem} ( \mu_A - \mu_A^{\varphi} ) - \xi_{\rm chem} \, , \\
   -\partial_t(f^2\partial_t\varphi) + \nabla\cdot(f^2\nabla\varphi) - f^2 m^2\varphi =& - \Gamma_{\rm chem} ( \mu_A^\varphi - \mu_A ) + \xi_{\rm chem} \, ,
\end{align}
with $\mu_A^{\varphi}=-\partial_t \varphi$.
After using lower order equations of motion, the stochastic equation for the pion field reads (see Appendix~\ref{pioneomapp})
\begin{equation} 
\label{wave_eq}
\chi_A\partial_t^2\varphi- f^2\partial_i^2 \varphi+f^2 m^2  \varphi - \lambda_A \nabla^2 \partial_t \varphi + \lambda_{m} m^2 \, \partial_t \varphi =  \xi, 
\end{equation}
where $\lambda_A$ and $\lambda_{m}$ are related to the coefficients described above (with $\Gamma_{\rm chem}\equiv 1/\zeta^{(2)}$):
\begin{align}
   \lambda_{A} &\equiv \sigma_A +  (\chi_A^{\rm nrm} v)^2 \,\zeta^{(2)} \, ,\\
\lambda_{m} &\equiv   (\chi_A^{\rm nrm} v)^2\, \zeta^{(2)} \, ,
\end{align}
and the noise $\xi$  satisfies 
 the fluctuation-dissipation relation:
\st
\llangle \xi(x) \xi(y) \rrangle = 2 T (-\lambda_A \nabla^2 + \lambda_{m}m^2 )\,  \delta(x-y) \, .
\stp


The equation of motion can be used to evaluate the corresponding propagators.
We will need the symmetrized correlation function
\begin{align}
   G_{ab,\rm sym}^{\varphi \varphi}(\omega,\q) \equiv \int d^4 x \, e^{i \omega t - i \q \cdot {\bm x} } \llangle \varphi_a(t,{\bm x}) \varphi_b(0) \rrangle  \equiv \delta_{ab} G_{\rm sym}^{\varphi\varphi}.
\end{align}
The retarded response
 function associated with the left hand side of \eqref{wave_eq} is  
\begin{align}
G_{R}^{\varphi\varphi}(\omega,\q) &=   \frac{1}{\chi_A} \frac{1}{-\omega^2 + \omega_q^2 -i \omega \Gamma_q},\\
\omega_q^2 &\equiv v^2(  q^2 + m^2),
\end{align}
where the pion velocity is given by $v^2=\frac{f^2}{\chi_A}$
and the attenuation is defined as
\begin{align}
\Gamma_{q} &\equiv D_A q^2 + D_m m^2, \qquad \text{where} \qquad D_A \equiv \frac{\lambda_A  }{\chi_A}, \qquad \text{and} \qquad D_{m} \equiv \frac{ \lambda_{m}}{\chi_A}.
\end{align}
$D_A$ and $D_{m}$ are the axial charge diffusion and damping coefficients, respectively.
Using the fluctuation-dissipation relation, or equivalently by solving \eqref{wave_eq} with
the noise, 
 we can obtain the symmetrized correlation function,
\begin{align}
   G_{\rm sym}^{\varphi \varphi} &=   \frac{1}{\chi_A} \frac{ 2 T \Gamma_q} {   (-\omega^2 + \omega_q^2)^2 + (\Gamma_q \omega)^2} \, .
\end{align}

The propagator is sharply peaked near the poles leading to an approximate expression
\st
G_{\rm sym}^{\varphi\varphi} \simeq \frac{ T}{2 \chi_A \omega_q^2} \left[ \rho(\omega,\omega_q) + \rho(\omega,-\omega_q)\right],
\stp
where
\st
\rho =  \frac{\Gamma_q  }{(-\omega + \omega_q)^2  + (\Gamma_{q}/2)^2 }.
\stp
This approximation is justified when the imaginary part of the dispersion relation, $\Gamma_q$, is negligible with respect to its real part $\omega_q$, i.e. $\Gamma_q/ \omega_q \ll 1$. 


\section{Dependence of the transport coefficients of QCD on the pion mass}\label{kubo}

Here we will first use the Kubo formula to deduce the dependence of the shear viscosity, $\eta$, bulk viscosity, $\zeta$, and iso-vector conductivity, $\sigma_I$, on 
the pion mass. The technical step is to integrate out the pion loop shown below to determine the corresponding fluctuations in the stress tensor or current.
\begin{figure}[h!]
\begin{center}
    \includegraphics[width=0.30\textwidth]{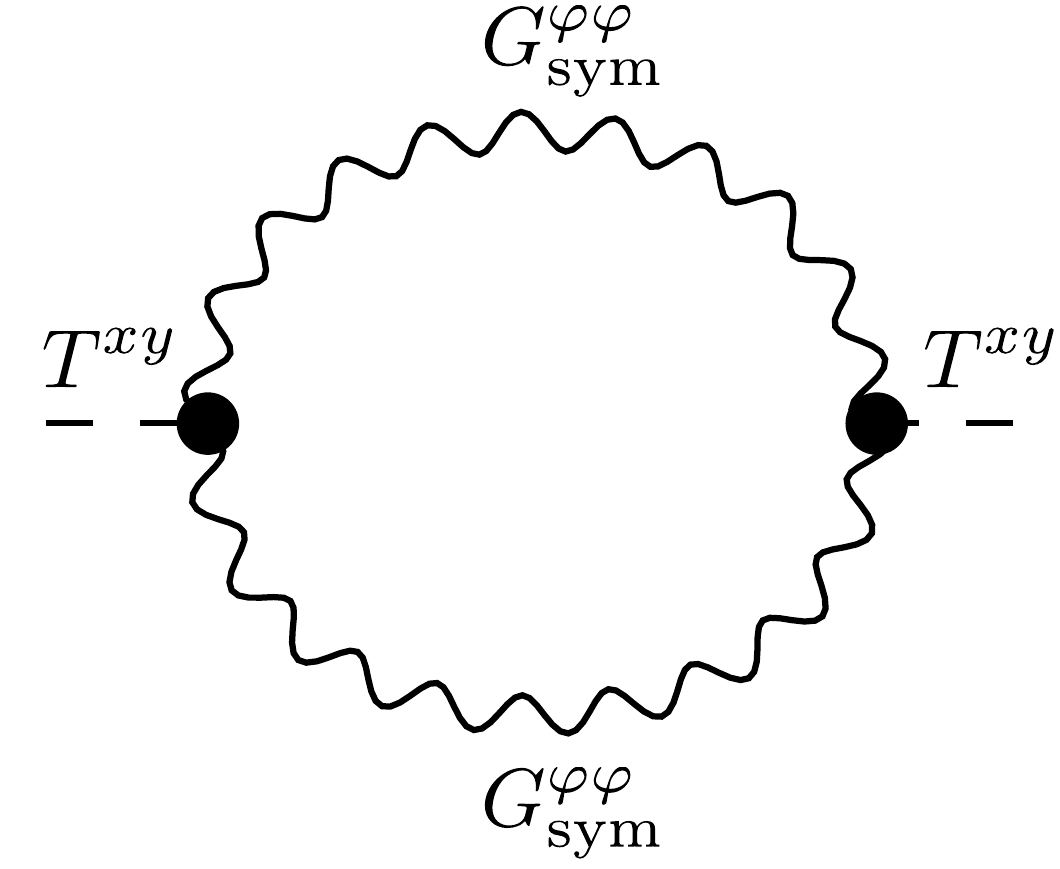}
    \caption{An example of a  hydrodynamic pion loop in Kubo furmulae.}
\end{center} 
\end{figure}

An equivalent approach to superfluid hydrodynamic loops 
is to develop a hydro-kinetic equation for the soft pions~\cite{Akamatsu:2016llw,An:2019osr}.
In this case the phase space distribution of soft pions  evolves according to a Boltzmann equation
with the normal fluid driving the distribution function out of equilibrium. The collision kernel of Boltzmann equation is determined by the axial charge diffusion and damping coefficients of the superfluid, $D_A$ and 
$D_{m}$, respectively.
A distinct advantage of the hydro-kinetic approach is that it can be simulated in expanding
environments, capturing the physics associated with the chiral fluctuations.
We will describe  this approach in \Sect{sec:pionkinetics} after analyzing the hydrodynamic loop.

\subsection{Kubo Formulae} 

The three transport coefficients of interest here are expressed as~\cite{forster1995hydrodynamic}
\begin{align}
    2 T \eta    =&   \int d^4x 
   \llangle \half \{T^{xy} (t,\x), T^{xy} 
   (0, {\bm 0})  \} 
   \rrangle, \\
    2 T \sigma_{I}  =&  \int d^4x 
   \frac{1}{d_A}  \llangle \half   \{  J_{V,a}^x (t,\x), J_{V,a}^x (0, {\bm 0})  \}
   \rrangle,  \\
    2 T \zeta =&  \int d^4x 
   \llangle \half \left\{ {\mathcal O}_{\rm bulk}(t,\x), {\mathcal O}_{\rm bulk}  
   (0, {\bm 0})  \right\} \rrangle.
\end{align}
Here $d_A=3$ is the dimension of the adjoint and we are summing over the isospin index. In determining the bulk 
viscosity is very convenient to use the operator 
\begin{align}
   \mathcal O_{\rm bulk} \equiv & c_s^2 T^{0}_{\phantom{0}0} + \frac{1}{3} T^{i}_{\phantom{i}i},
\end{align}
where $c_s^2$ is a fixed parameter for the system at temperature $T$.
This operator has several related advantages over $T^{\mu}_{\;\mu}$~\cite{Jeon:1995zm}. Specifically,
it is invariant in equilibrium under 
small shifts in the temperature, and therefore it is not necessary to 
impose the Landau matching condition when perturbing the system.  It also behaves smoothly as the spatial momentum $\k \rightarrow 0$, while $T^{\mu}_{\;\mu}$ does not~\cite{Romatschke:2009ng,Hong:2010at}.

To evaluate the pion contributions to these correlations we will need the symmetrized correlation function
discussed in the previous section and more explicit expressions for the operators of interest.
Here we have
\begin{align}
   T^{xy}=&  f^2  \partial^x\varphi_{a} \partial^{y} \varphi_a, \\ 
   J_{V,a}^{x} =&  -f^2 f_{abc} \partial^x\varphi_{b} \varphi_c,  \\
   {\mathcal O}_{\rm bulk} =&   
  \left[ p_{\varphi} + 
    \frac{1}{3} f^2(\nabla \varphi_a)^2  - c_s^2 \left( \chih (\partial_t \varphi_a)^2  -  \frac{\partial (\beta p_{\varphi}) }{\partial \beta} \right)  \right],  \\
  p_{\varphi} =& \frac{1}{2} \chi_A \left[ (\partial_t\varphi_a)^2 - v^2 (\nabla \varphi_a)^2 - v^2 m^2 \varphi_a^2 \right] .
\end{align}
Evaluating the Feynman graphs for the shear stress, current, and bulk operator gives
\begin{align}
   2 T \eta  &=  2 T \eta^{(0)}(\Lambda) + 2 d_A f^4 \int^{\Lambda} \frac{dq^0 d^3q}{(2\pi)^4} \,    (q^x q^y)^2  \, (G^{\varphi \varphi}_{\rm sym}(q^0,q) )^2, \\
   2 T \sigma_I  &=   2T \sigma_I^{(0)}(\Lambda) + 2  T_{A}f^4 \int^{\Lambda} \frac{dq^0 d^3q}{(2\pi)^4} \, (q^x)^2  \, (G^{\varphi\varphi}_{\rm sym}(q^0, q))^2, \\
   2 T \zeta &= 2 T \zeta^{(0)}(\Lambda) + 2d_A  \int^{\Lambda}  \frac{dq^0 d^3q}{(2\pi)^4} \, (\mathcal N_{\rm bulk}(q_0,q))^2 \, (G^{\varphi\varphi}_{\rm sym}(q^0,q))^2,
\end{align}
where $T_{A} = 2$ is the trace of the adjoint.  The numerator algebra associated with the 
operator $\Obulk$ evaluates to
\st
{\mathcal N}_{\rm bulk} =  \half \left(\chih + c_s^2 \frac{\partial(\beta \chih)}{\partial \beta}\right)  \, (q_0^2  -\omega_q^2)
+ \frac{\chih}{3} v^2 q^2 - c_s^2 \chih \left(q_0^2 +  \frac{\beta}{2} \frac{\partial \omega_q^2 }{\partial \beta} \right).
\stp
For each integral we will  perform the $q^0$ integration first.  The propagators  are sharply peaked near
$q^{0} = \pm \omega_q$, and cross terms in $(G^{\varphi\varphi}_{\rm sym})^2$ can be neglected in the integration. Performing
the $q_0$ integral we find
\begin{subequations}
   \label{transport-eqs}
\begin{align}
   \eta  &=\eta^{(0)}(\Lambda) +  d_A \int^{\Lambda} \frac{d^3q}{(2\pi)^3}    \left(\frac{\partial \omega_q}{\partial q_x}  q_y \right)^2  \left( \frac{T}{\omega_q^2} \right) \frac{1}{\Gamma_q},  \\
   \sigma_I  &=\sigma_I^{(0)}(\Lambda) + T_{A} \int \frac{ d^3q}{(2\pi)^3} \left( \frac{\partial \omega_q}{\partial q_x} \right)^2  \left(\frac{T}{\omega_q^2 } \right) \frac{1}{\Gamma_q}, \\
   \zeta &=\zeta^{(0)}(\Lambda)+ d_A  \int^{\Lambda} \frac{d^3q}{(2\pi)^3} \left[ \frac{\q}{3} \cdot \frac{\partial \omq}{\partial \q} - c^2_s \frac{\partial (\beta\omq) }{\partial \beta} \right]^2 \left(\frac{T}{\omega_q^2} \right)\frac{1}{\Gamma_q}.
  \end{align}
\end{subequations}
As is briefly described in the next subsection these expressions are familiar from kinetic theory.

These integrals depend on the two transport coefficients, $D_A$ and $D_{m}$, the thermodynamic properties
of the soft pions, $m_p^2 =v^2 m^2$ and $v^2 = f^2/\chi_A$, as well as the speed of sound squared, $c_s^2$, 
which can be determined from the Euclidean measurements. These quantities enter in
the final results as
\begin{align}
   \tilde v^2 &= v^2 - \frac{T}{2} \frac{\partial v^2}{\partial T},  \\
   \tilde m_p^2 &= m_p^2 - \frac{T}{2} \frac{\partial m^2_p }{\partial T} , 
\end{align}
and via a dimensionless ratio
\st
r =  \sqrt{ \frac{D_{m}}{D_A} } .  
\stp
It is worthwhile to point out that in chiral perturbation theory $r= \sqrt{4/3}$~\cite{derek_tbp}. 

Evaluating the integrals, we find the final expressions to the corrections of the transport coefficients:
\begin{subequations}
         \label{finaltransport}
      \begin{align}
         \zeta =&  \zeta^{(0)}_{\rm phys} + 
\frac{d_A T m}{8\pi D_A }  \left[ \left( \frac{c_s^2}{1+r} \, \frac{\tilde m^2_p}{m^2_p} - \frac{1 + 2r}{1+r}
\left(\tfrac{1}{3}  - c_s^2 \frac{\tilde v^2}{v^2}   \right) \right)^2 -
 (4 +2r) \left(\tfrac{1}{3}  - c_s^2 \frac{\tilde
v^2}{v^2}   \right)^2 \right], \\        
\eta  =& \eta^{(0)}_{\rm phys} -\frac{d_A T m}{120 \pi D_A}  \left[ \frac{ 2 r^3+4 r^2+6 r+3}{ (1+r)^2} \right],  \label{finaltransport2} \\         
\sigma_{I} =&  (\sigma_{I})^{(0)}_{\rm phys}+  \frac{T_A T}{24 \pi m D_A} \left[   \frac{1+ 2r}{(1 + r)^2 } \right]  .\label{finaltransport3}
      \end{align} 
   \end{subequations}
Here  the shear viscosity and the bulk viscosity are renormalized quantities
\begin{subequations}
\begin{align}
\zeta^{(0)}_{\rm phys}=& \zeta^{(0)}(\Lambda) +  \frac{ d_A T \Lambda}{2\pi^2 D_A } \left(\tfrac{1}{3} - c_s^2 \frac{\tilde v^2}{v^2} \right)^2,   \\
   \eta^{(0)}_{\rm phys}=& \eta^{(0)}(\Lambda) +  \frac{ d_A T \Lambda}{30 \pi^2 D_A }. 
\end{align}
\end{subequations}
In each case, the  ``zero'' transport coefficients (e.g. $\zeta^{(0)}$) are the parameters in the chiral limit, $m_q=0$.
The conductivity $\sigma_I$ is not renormalized,
and  its soft pion contribution is proportional to  the inverse screening mass,
$m^{-1}$. This contribution diverges in the chiral limit and is parametrically
larger than $(\sigma_I)_{\rm phys}^{(0)}$. This reflects the  fact that in this
limit the soft pion is a free particle which transports isospin.

As emphasized in 
\cite{Son:2002ci} many of the parameters in \eqref{finaltransport} can be evaluated on the lattice~\cite{Brandt:2014qqa}.  
Indeed all of the parameters of the ideal superfluid hydro, such 
as $v^2, c_s^2, m_p^2$ and $m^2$, are amenable to a Euclidean computation, while the viscous parameters $D_A$ and $r$ must be extracted from data or estimated from theoretical considerations. 
We will analyze the behavior of \Eq{finaltransport} near the $O(4)$ critical point in \Sect{discussion}.

\subsection{Kinetic approach}
\label{sec:pionkinetics}

The physical content of hydrodynamic loop calculations, such as described
in the previous section, are  (always) equivalent to 
deriving a Boltzmann equation for the hard sound modes in the plasma and using
this Boltzmann equation  to analyze the response~\cite{Akamatsu:2016llw,An:2019osr}.
Indeed, our results for the transport coefficients \eqref{finaltransport} are much more transparently obtained from a hydro-kinetic Boltzmann equation for the soft pion phase-space distribution function $f_\pi(x,q_i)$,
which takes the form of a relaxation time like approximation. 
We are  motivated by a similar Boltzmann equation for sound modes in normal hydrodynamics~\cite{An:2019osr}. 

First we generalize the linear analysis of the previous section to a flowing fluid background in Appendix~\ref{pioneomapp},  using
the scale separation depicted in \Fig{cartoon}. The ideal terms in the equation
of motion 
are of order $\partial^2 \varphi \sim m^2 \varphi$, while flow
corrections to these terms  are of order $(\partial \varphi) (\partial u) \sim m \varphi /L$. Denoting the mean free path $\lambda$, the dissipative terms in the equation of motion are of order $ \lambda\partial^3 \varphi \sim \lambda m^3 \varphi$, while flow corrections to these terms are of order $(\lambda/L) m^2$ and are ignored.  We are thus working in a kinetic regime where
\st
(\lambda/L) m^2  \ll   (m/L) \sim  \lambda m^3 \ll m^2   \, .
\stp
With these approximations the stochastic wave equation takes the form
\st
-\partial_{\mu} (\chih \, G^{\mu\nu} \partial_{\nu} \varphi) + f^2 m^2 \varphi    -\lambda_{A}  \nabla_\perp^2 \partial_{\tau} \varphi  + \lambda_m  m^2 \partial_\tau \varphi = \xi \, .
\stp
Here $\nabla_\perp^{\mu}\equiv\Delta^{\mu\nu}\partial_{\nu}$ and $\partial_{\tau} = u^{\mu} \partial_{\mu}$ are the local spatial and temporal  derivatives,
and the pion field moves in an effective metric
created by the flowing fluid
\st
\label{eq:fluidmetric}
G^{\mu\nu}(x) \equiv -u^{\mu}(x) u^{\nu}(x) + v^2(x) \Delta^{\mu\nu}(x) \, ,
\stp
where  $v^2(x)\equiv f^2/\chi_A$ is the local pion velocity.

Appendix~\ref{app:kin} shows that 
under the evolution of the stochastic  wave equation,
the pion phase-space distribution $f_\pi(x,q_i)$ evolves
according to a Boltzmann equation which takes the form
\begin{equation}
   \label{transporteqnfinal}
\frac{\partial \mathcal H}{\partial q_\mu}\,    \frac{\partial f_\pi}{\partial x^{\mu} }   
   -  \frac{\partial \mathcal H} {\partial x^i}
   	\, \frac{\partial f_\pi  }{\partial q_i} = - \Gamma_q  \left[\omega_q \, f_\pi - T \right] \, , 
\end{equation}
where the effective Hamiltonian,
\st
\mathcal H(x,q) = \frac{1}{2} G^{\mu \nu}(x) q_{\mu} q_{\nu}  +  \frac{1}{2} v^2(x) m^2(x)  \, ,
\stp
is a function of the four vectors $x$ and $q$.
Given the covariant momenta $q_i$, the covariant energy component $q_0$ is
found by solving the on-shell constraint, $\mathcal H(x,q) = 0$, taking the negative root $q_0 = -h_+(x,q_i)$, see
\eqref{hpmonshell}. The components 
$G^{\mu\nu} q_{\nu}= \partial \mathcal H/\partial q_{\mu}$ should be distinguished from $q^{\mu} \equiv \eta^{\mu\nu} q_{\nu}$. 
The
damping rate $\Gamma_q$ and dispersion curve $\omega_q=-u^{\mu} q_{\mu}$ are to be evaluated in
the rest frame of the fluid, see \eqref{RetardedG}. The equilibrium distribution is simply the classical part of the Bose-Einstein distribution function, $T/\omega_q$.  We note
that the Boltzmann equation can also be written 
\begin{equation}
   \label{transporteqnfinalb}
   \frac{\partial \mathcal H}{\partial q_0}\,  \left[   \frac{\partial f_\pi}{\partial t} 
      + \frac{\partial h_+}{\partial q_i} \frac{\partial f_\pi}{\partial x^i} 
   - \frac{\partial h_+}{\partial x^i} \frac{\partial f_\pi}{\partial q_i}  \right]
      = - \Gamma_q  \left[ \omega_q  \, f_\pi - T \right] \, .
\end{equation}

Once the phase space distribution is found, the pion contribution to the stress tensor is given by the superfluid stress in \eqref{Tmunudef}. Recalling
that the phase space distribution $f_\pi(x,q_i)$ is the Wigner transform of the noise averaged two point function  $\llangle \varphi(x) \varphi(y) \rrangle$, \app{BoltzStress} shows that pion contribution to average stress tensor evaluates to
\begin{equation}
   \label{Tmunuboltz}
  T^{\mu\nu}_\pi = d_A \int \frac{d^3q_i}{(2\pi)^3 (\partial \mathcal H/\partial q_0 ) } \left[ \omega_q \frac{\partial (\beta \omega_q) }{\partial \beta} u^{\mu} u^{\nu}  +  v^2 \Delta^{\mu\alpha} \Delta^{\nu\beta} q_{\alpha} q_{\beta} \right] f_{\pi} (x,q_i)  \, .
\end{equation}

Given this Boltzmann equation for the soft pion distribution and the stress tensor,  
familiar steps from the relaxation time approximation
lead to the shear and bulk viscosities  presented in \eqref{transport-eqs} with relaxation time $1/\Gamma_q$ and dispersion curve $\omega_q$.
We have not derived the isospin conductivity using the kinetic approach in this paper.

\section{Discussion and behavior near the chiral critical point}\label{discussion}

In this section we will estimate our results for the transport coefficients, \Eq{finaltransport}, near the $O(4)$ critical point.
\subsection{The chiral phase transition: a brief review}
First we review the expected scaling behavior of various quantities 
following \cite{Son:2001ff,Rajagopal:1992qz}. 
The order parameter, $ \sigma(x) \equiv \bar\psi \psi(x) $, and the inverse correlation length, $m_\sigma$,
have the following scaling behavior near the critical point
\begin{equation}\label{eq:def_critical_exponent}
\langle \bar\psi \psi \rangle \sim t^\beta,\quad m_\sigma\sim t^{\nu},
\end{equation}
where the reduced temperature is $t=|T-T_c|/T_c$, and $\beta$ and $\nu$ are the
usual critical exponents\footnote{We are only interested in temperatures
below $T_c$ in this study.}.
In the vicinity of the critical point, the static correlation function of the order parameter behaves like
\begin{equation}\label{order_parameter}
   \int d^3 x\;  e^{-i\x\cdot \q }\, \langle \sigma({\bm x}) \sigma(0) \rangle \sim \frac{T}{|\q|^{2-\eta}},
\end{equation} 
for momentum, $|\q|$, much larger than $m_\sigma$, but smaller than the temperature $T$, $m_\sigma\ll|\q| \ll T$. 
The critical exponent $\eta$ is small in practice, and can be related to  $\beta$ and $\nu$ using the hyperscaling relation
\begin{equation}
2\beta = \nu \left( d -2+ \eta\right),
\end{equation}
where $d=3$ is the spatial dimension. Finally, 
the order parameter relaxation rate scales with correlation length as~\cite{Rajagopal:1992qz} 
\begin{equation}
   \Gamma_\sigma \sim (m_\sigma)^{z}, 
\end{equation} 
where $z=d/2$ is the dynamical critical exponent.
As we will describe in the next paragraphs,  $\Gamma_\sigma$ is of order $D_A m_\sigma^2$, and therefore we will define $\Gamma_{\sigma} \equiv D_A m_\sigma^2$ 
in our estimates below.
A summary of the relevant scalings can be found in Table 1.
\begin{table}
\centering
\begingroup
\setlength{\tabcolsep}{5pt}
\def\arraystretch{1.2}
\begin{tabular}{ |c|c|c|c| }
 \hline
 physical quantity & symbol& scaling & estimate \\ [0.5ex] 
 \hline\hline  
 order parameter &$ \llangle \sigma \rrangle =\langle\bar\psi \psi\rangle$ & $t^\beta$  & $\beta\simeq 0.380 $ \\ 
 inverse correlation length & $m_\sigma$ & $t^\nu$  &   $\nu \simeq 0.738$ \\ 
 static correlation function & $\int d^3 x\;  e^{-i\x\cdot \q } \,\langle \sigma(\x) \sigma(0) \rangle$ & $T |\q|^{\eta-2}$ & $\eta\simeq 0.03$  \\
 $\sigma$ relaxation rate & $ \Gamma_\sigma \equiv D_A m_\sigma^2 $ & $m_\sigma^{z}$ & $z=\frac{d}{2}$ \\  [0.5ex]
 \hline
 axial susceptibility &  $\chi_A$  & $\chi_0$ & {\rm const} \\
 $(\mbox{pion velocity})^2$ &  $v^2 {=}f^2/\chi_A$  &   $t^{\nu(d-2) }$ &   $t^{0.738}$ \\
 $(\mbox{screening mass})^2$ &  $m^2$  & $m_q   t^{\beta - (d-2)\nu }$ &  $m_q\, t^{-0.358}$ \\
 $(\mbox{pole mass})^2$ & $m^2_p{=}v^2m^2$  &   $m_q t^{\beta }$ &  $m_q \,t^{0.380}$ \\ [0.5ex] \hline 
\end{tabular}
\endgroup
\\
\caption{\label{O4table} Here we list the critical scaling of relevant parameters near the chiral critical point, as discussed in \cite{Son:2001ff}. }
\end{table}

In the chirally broken phase close to $T_c$, but not so close that fluctuations
in $\sigma$ are important,  the Lagrangian in \Eq{superfluidlagrange} applies.
The static pion propagator $\pi_a = \llangle \bar\psi\psi \rrangle  \varphi_a$ 
at momentum scale $m\ll |\q| \ll m_\sigma $
can be read off from the Euclidean version of the Lagrangian
\begin{equation}
   \label{pion_propagator}
   \int d^3 x\; e^{-i\x\cdot \q }\, \langle \pi_a(\x)  \pi_b(0) \rangle\approx \delta_{ab} \frac{ T  \langle \bar\psi\psi \rangle^2}{f^2} \frac{1}{|\q|^2}.
\end{equation}
In the vicinity of the phase transition,
 the $\pi$ and $\sigma$ propagators become degenerate.
 Thus, the scaling of the two propagators (\Eq{order_parameter} and \Eq{pion_propagator}) must be the same 
at their boundaries of applicability  $|\q|\sim m_\sigma$, leading to a relation between the pion decay constant $f^2$ , $m_\sigma^2$, and $\langle \bar\psi \psi\rangle $, 
\begin{equation}
%
%
f^2\sim m_\sigma^{-\eta} \langle \bar\psi \psi\rangle^2\sim t^{\nu(d-2)}.
\end{equation}
Then the pion velocity near $T_c$ scales like
\begin{equation}
v^2= \frac{f^2}{\chi_A} \sim  t^{\nu(d-2)}, 
\end{equation}
where $\chi_A \simeq \chi_0$  is approximately constant near $T_c$.
The screening mass, $m$, can be related to the pion decay  constant $f^2$, the condensate $\llangle \bar\psi\psi\rrangle$, and the quark mass $m_q$ via
\begin{equation}\label{screening-mass}
m^2= - m_q \frac{\langle\bar \psi \psi  \rangle }{f^2} \sim m_q t^{\beta-(d-2)\nu}.
\end{equation}
Similarly, the pole mass scales like
\begin{equation}
   m_p^2= v^2 m^2 = -m_q \frac{\langle\bar \psi \psi  \rangle}{\chi_\Aa} \sim m_q t^{\beta} \,.
\end{equation}

Close to $T_c$, but again not so close that $\sigma$ fluctuates, the 
hydrodynamic analysis applies and
the pion dispersion curve for $ q \gg m $ reads
\begin{equation}
   \omega(q)= v |\q  | - \tfrac{i}{2} D_A q^2 \, .
\end{equation}
As we approach the phase transition, the real and imaginary parts of
the dispersion curve become the same order of magnitude. Also,  the pion and $\sigma$  damping rates should scale similarly near $T_c$ at the  boundaries of applicability $|\q|  \sim m_\sigma$. This reasoning yields the following 
 estimates
\begin{equation}
   v m_\sigma\sim  D_A m_{\sigma}^2 \sim (m_\sigma)^z \, .  
\end{equation}
For definiteness we will define the relaxation rate $\Gamma_{\sigma}$ using the axial charge diffusion coefficient $\Gamma_\sigma \equiv D_A m_\sigma^2$. The  temperature scaling of $D_A$ is
%
%
\begin{equation}\label{Da-scaling}
D_A\sim t^{\nu(\frac{d}{2}-2)}= t^{-\nu/2}.
\end{equation}


\subsection{The transport coefficients near $T_c$}
Armed with these scaling relations, we can determine the temperature
and quark mass 
dependence of the transport coefficients near $T_c$.  We are assuming
that we are not so close to $T_c$ that the $\sigma$ field fluctuates strongly.
The $\sigma$ fluctuations cannot be neglected when the screening
mass in  \eqref{screening-mass} becomes of order  $m_\sigma$, which yields 
\st 
\label{limit-scaling}
\frac{m}{m_\sigma}  \sim \sqrt{m_q}\,  t^{(\beta - 3\nu)/2}  \, ,
\stp
and therefore the analysis breaks down when $t \sim (\sqrt{m_q})^{1.091}$.  Since the pion mass $m_\pi \propto \sqrt{m_q}$ is fairly massive compared to, e.g. $2\pi T_{c}$, 
it is likely that $\sigma$ fluctuations
can never be completely ignored over the temperature range relevant to heavy ion collisions.


Nevertheless, we wish to evaluate the temperature and 
quark mass dependence of the corrections to the transport 
coefficients given in \Eq{finaltransport}.
Using the scalings described above, first note that 
\begin{align}
   \frac{\tilde m_p^2 }{m_p^2} =& \frac{\beta}{2t}  \qquad \beta = 0.380, \\
   \frac{\tilde v^2 }{v^2} =&  \frac{\nu}{2t}  \qquad \nu = 0.737. 
\end{align}
near the critical point.
In addition, we make the approximation $\nu \simeq 2\beta$, and note that 
the speed of sound remains finite $c_s^2 \simeq c_{s0}^2$ in the $O(4)$ model~\cite{Engels:2011km}. Thus, the bulk viscosity in \eqref{finaltransport} reduces  to
\st
\label{zetaTc}
\zeta = \zeta^{(0)} - \frac{d_A T m}{8\pi D_A} \left(\frac{\beta c_{s0}^2 }{t}\right)^2 \left[ 
 \frac{8 r^3+16 r^2+16 r+7}{4 (1+r)^2 }
\right] ,
\stp
near the critical point.
The parameter $r$ approaches an order one constant near the critical point~\cite{Rajagopal:1992qz,fanglida_tbp}, and thus 
our results for $\zeta$, $\eta$, and $\sigma_I$  depend on 
an unknown constant\footnote{Translating the notation of the current work
into the notation  of the original
reference~\cite{Rajagopal:1992qz,fanglida_tbp},  we have $r^2 =\Gamma/(\Gamma
+ \gamma/\chi)$ where $\Gamma$ and $\gamma$ are the two dissipative
parameters characterizing the $O(4)$ Langevin model of
\cite{Rajagopal:1992qz}. These parameters scale similarly near near critical
point, $\Gamma m_\sigma^2 \sim (\gamma/\chi) m_\sigma^2 \sim  m_\sigma^{z}$,
and thus $r$ is approximately constant.
}.  
However, we have found that the $r$ dependence of \eqref{finaltransport}
and \eqref{zetaTc} is mild, and changing $r$ from zero to one
changes the shear and conductivity coefficients by  less than 25\%, and the bulk coefficient by 60\%. 
(In low temperature chiral perturbation theory $r=\sqrt{4/3}$, see \cite{derek_tbp}).
Thus, for simplicity, we will set $r=0$ below, and estimate a constant factor of two uncertainty from this ansatz.

With these rough approximations the three transport equations read
\begin{subequations}
   \label{viscosity}
\begin{align}
   \zeta &= \zeta^{(0)}_{\rm phys}  -
   \frac{21}{32\pi } \left[\frac{Tm^3}{D_A m^2} 
   \left( \frac{\beta c_{s0}^2}{t} \right)^2 \right] , \\
    \eta &=  \eta^{(0)}_{\rm phys}  
    -   \frac{3}{40\pi} \left[ \frac{  T m^3}{ D_A m^2} \right] ,  \\
    \label{sigmaIfinal}
    \sigma_I &=   (\sigma_{I})_{\rm phys}^{(0) } + \frac{1}{12\pi} \left[ \frac{T m}{ D_A m^2 }  \right]  \, .
\end{align}
\end{subequations}
In each case, the  ``zero'' transport coefficients (e.g. $\zeta^{(0)}$) are the coefficients in the chiral limit $m_q=0$, and the additional bits describe how these parameters
depend (non-analytically) on the quark mass. 
The soft pion  parts
can be easily understood as the pion contribution to the corresponding susceptibility\footnote{For instance, the soft pion contribution to the isospin susceptibility is 
   \st
   (\chi_I)_{\pi}^{\rm soft} \sim \int_p \frac{d^3p}{(2\pi)^3} \left. \frac{\partial n_p}{\partial \mu_I}\right|_{\mu=0} \sim T m \, ,
   \stp
   where $n_p \simeq T/(\omega_p-\mu_I)$. Similarly, the soft pion  
   enthalpy (the susceptibility associated with $\eta$) is $(e+p)_\pi^{\rm soft} \sim Tm^3$.
} (the numerators in \eqref{viscosity})
divided by the damping rate, $\Gamma_q \sim D_A m^2$.

The isospin (or charge) conductivity is dominated by the soft pion contribution, which diverges in the chiral limit  and therefore is large compared
to $(\sigma_{I})^{(0)}_{\rm phys}$ until $m \sim m_{\sigma}$.  This reflects the fact 
that on a length scale $m^{-1}$, the Goldstone bosons can transport charge freely rather than diffusively.
When $m \sim m_{\sigma}$, 
the iso-vector conductivity \eqref{sigmaIfinal} and the axial charge diffusion coefficient \eqref{Da-scaling} have the same scaling with reduced
temperature
\begin{equation}
\sigma_I \sim  \frac{T m_\sigma}{ \Gamma_\sigma }   
 \sim t^{-\nu/2} \sim \chi_A D_A \, ,
\end{equation}
as is required by the restoration of vector-axial-vector symmetry  at the critical point.

Finally, we may put in the expected scaling for $\Gamma_{\sigma}$, $m_\sigma$, and $m/m_\sigma$ to find:
\begin{subequations}
   \label{asymptotic}
\begin{align}
   \Delta \zeta =&   - C_{\zeta}  \, \sqrt{m_q} \, t^{\beta/2}\left( \frac{\beta c_{s0}^2}{t} \right)^2  ,  &  \Delta \zeta \propto& \; -t^{-1.81},   \\
   \Delta \eta =& -C_{\eta} \, \sqrt{m_q}\,   t^{\beta/2},  & \Delta \eta \propto& \; -t^{0.19},  \\
   \sigma_I =&   C_{\sigma_I}   \,\frac{ t^{\nu - \beta/2}} {\sqrt{m_q}}, & \sigma_I \propto& \;  t^{0.548} \, .
\end{align}
\end{subequations}
Thus approaching $T_c$ from below, the pion contribution to the bulk viscosity decreases sharply, while the shear viscosity contribution grows mildly. 
The transport of soft pions dominates the isospin conductivity, and  the conductivity decreases as the  damping rate of the Goldstone mode  increases near $T_c$.
%

In practice, the asymptotic behavior in \eqref{asymptotic} will be difficult to see in the narrow window where the theory applies.
Indeed, we are only able to understand the modifications of the transport coefficients due to pions in the broken phase. As pointed out in \eqref{limit-scaling}, our results are valid up to a scale where the fluctuations of the order parameter $\sigma$ becomes large. A natural follow up would be to include such fluctuations, significantly increasing the range of applicability of the current study.

In addition to the modifying the transport coefficients of the fluid, chiral critical fluctuations modify the dispersion relation  of soft pions, e.g. $\omega_q^2 \simeq v^2 q^2 + m^2_p$ with $v$ and $m_p$ small compared to their vacuum values.  These modifications are expected to lead to an anomalous enhancement  of pions at small momenta \cite{Rajagopal:1992qz,Son:2001ff}, which is a phase space region ideally suited to the upcoming ITS detector in ALICE~\cite{ALICE}. Currently, there is some evidence for such a soft pion enhancement -- see for example Fig. 3 in \cite{Devetak:2019lsk} and Fig. 11 in \cite{Acharya:2019yoi}. In the future we hope to use the kinetic equations developed in \Sect{sec:pionkinetics} to quantitatively compute these enhancements and their associated fluctuations.


\begin{acknowledgments}
We thank Yukinao Akamatsu and Juan Torres-Rincon for collaboration during the initial stages of this project.  We are grateful to Mauricio Martinez, Paolo Glorioso, Misha Stephanov, and Rob Pisarski for helpful conversations. 
This work is supported by 
the U.S. Department of Energy, Office of Science, Office of Nuclear Physics,
grants Nos.
DE\nobreakdash-FG\nobreakdash-02\nobreakdash-08ER41450. AS is supported by the Austrian Science Fund (FWF), project no. J4406.
\end{acknowledgments}

\begin{appendix}

\section{Entropy production}
\label{app:ent}

In this appendix, we describe the computation of the entropy production in detail, repeating formulas as necessary to keep the presentation self-contained. The entropy is 
\st
s_{U} = \frac{\varepsilon_U + p_U - \mu_L \cdot n_L - \mu_R \cdot n_R }{T} \, .
\stp
The thermodynamic relation is a consequence of the independent variables used to describe the partition function
\st \label{thermo-relation}
dp_U= s_U d T  + n_L \cdot d\mu_L + n_R \cdot d\mu_R  
- \frac{f^2}{8}  dL^2 + 
\frac{f^2 m^2}{8} d( U \cdot \mathcal{M}^\dagger +\mathcal{M}\cdot U^{\dagger}).
\stp
The ``extra'' superfluid differentials at fixed $T$ and $\mu$ follow from the form of the action \eqref{superfluidlagrange},
and the discussion surrounding the derivation of the stress tensor \eqref{Tmunudef}. 
Using  $u^{\mu} L_{\mu} = -i dU U^{\dagger}$,
they  can be written
\st
(dp_{U})_{T,\mu} \equiv
 - \frac{f^2}{8} dL^2  + i\frac{f^2 m^2}{8 } u^{\nu} L_{\nu} \, \cdot  (U  \mathcal{M}^\dagger   - \mathcal{M}  U^\dagger). 
\stp
Then the entropy current satisfies
\st
\partial_{\mu} (s_U u^{\mu}) = ds_U + s_U \partial u ,\nonumber
\stp
where we have implemented the following shorthand
\st
u^{\mu} \partial_{\mu}s_U   = ds_U, \qquad \partial u \equiv \partial_{\mu} u^{\mu} \, ,
\stp
and note that the differentials in \eqref{thermo-relation} can be interpreted 
with an analogous notation, e.g.~$dT = u^{\mu} \partial_{\mu} T$. Inserting
the definition  definition of entropy yields 
\begin{multline}
   \label{entropyintermediate}
   \partial_\mu (s_U u^{\mu}) = \frac{1}{T} \left[d\varepsilon_U + (\varepsilon_U + p_U) \, \partial u\right]  +  
   \frac{1}{T}  (dp_{U})_{T,\mu}
 - \frac{\mu_L}{T}  \left[ dn_L  + n_L \,\partial u \right]  
 - \frac{\mu_R}{T}  \left[ dn_R  + n_R\, \partial u \right]  .
\end{multline}

Now we should use the equations of motion of energy conservation, $u_{\nu} \partial_{\mu} T^{\mu\nu}=0$,
and current partial conservation 
\begin{subequations}
\begin{align}
   \partial_\mu J^\mu_L =&- i \frac{f^2 m^{2}}{8 }(U \mathcal{M}^\dagger- \mathcal{M} U^\dagger), \\
   \partial_\mu J^\mu_R =&+ i \frac{f^2 m^{2}}{8 }(\mathcal{M}^\dagger U - U^\dagger \mathcal{M} ), 
\end{align}
\end{subequations}
to evaluate the terms in square brackets of \eqref{entropyintermediate}.
Note we have imposed the microscopically exact PCAC relation. 
From the body of the text, the stress tensor, currents, and Josephson relation, can be written
\begin{subequations}
\begin{align}
   T^{\mu\nu} =& (\varepsilon_U +p_U) u^{\mu} u^{\nu} + p g^{\mu\nu}+ \frac{f^2}{4}  L^{\mu}  \cdot L^{\nu} +  \Pi^{\mu\nu},  \\
   J^{\mu}_L  
   =& n_L u^{\mu} +  \frac{f^2}{4} L^{\mu} + q_L^{\mu} , \\
   J^{\mu}_R =&    
   n_R u^{\mu} + \frac{f^2}{4} R^{\mu} + q_R^{\mu}  ,  \\
    \label{eq:Josephsonapp}
   -\tfrac{1}{2} u^{\mu} L_{\mu} =& \mu_A + \mu_A^{\diss}  ,
\end{align}    
\end{subequations}
with dissipative strains $\Pi^{\mu\nu}$, $q_L^{\mu}$, $q_R^{\mu}$, and  $\mu_A^{\diss}$.


The left current partial conservation  equation yields
 \begin{align}\label{left_cc}
    -\mu_L\cdot (d n_L + \cdot n_L \partial u ) =&  
    \mu_L\cdot \partial_{\mu}  q_L^{\mu} 
  +  \mu_L\cdot \left[ \frac{1}{4} \partial_{\mu} (f^2 L^{\mu} )
    + i \frac{f^2 m^{2}}{8 } (U \mathcal{M}^\dagger- \mathcal{M} U^\dagger)   \right] , \nonumber \\
    \equiv& \mu_L \cdot \partial_{\mu}  q_L^{\mu}  + \tfrac{1}{2}\,  \mu_L \cdot \Theta_s  ,
 \end{align}
 where we have defined the superfluid expansion scalar discussed in the 
 text (see \eqref{eq:thetasviaS})
 \st
\Theta_s \equiv \left[ \partial_{\mu} \left( \frac{f^2}{2} L^{\mu} \right)  + \frac{ f^2 m^2}{4} \,i( U \mathcal M^\dagger - \mathcal M U^\dagger) \right]  \, . \nonumber
\stp
Similarly, the right current partial conservation  equation yields
 \begin{align}\label{right_cc}
-\mu_R\cdot (d n_R+n_R \partial u)  
=&  \mu_R\cdot \partial_{\mu}  q_R^{\mu}  
 + \mu_R\cdot  \left[\frac{1}{4} \partial_{\mu}  (f^2 R^{\mu} )
  - i \frac{f^2 m^{2}}{8 }  ( \mathcal{M}^\dagger U-  U^\dagger \mathcal{M}) \right], \nonumber \\
 =&  \mu_R\cdot \partial_{\mu}  q_R^{\mu}  
 - \tfrac{1}{2} U^\dagger \mu_R U \cdot \Theta_s .
 \end{align}
 In passing to the second line we have
used the definition of $R^\mu$ as $R^\mu= - U^\dagger L^{\mu}U $ and the definition of $L_{\mu}=-i \partial_{\mu}U U^\dagger$
to rewrite:
\begin{align}
\mu_R \cdot  \partial_{\mu}  (f^2 R^{\mu} )
&=- U\mu_R  U^\dagger \cdot \partial_{\mu}  (f^2   L^{\mu}   )  \, .
\end{align}

Next, we consider the timelike projection of the energy momentum tensor conservation equation, $ -u_{\nu} \partial_{\mu} T^{\mu\nu}=0$:
\begin{align} 
%
   d\varepsilon_U  + (\varepsilon_U + p_U) \partial u =& u_{\nu}\,  \frac14 \partial_{\mu} (f^2 L^{\mu} \cdot L^{\nu}) +  u_{\nu} \partial_{\mu} \Pi^{\mu \nu}, \nonumber \\
  =& \frac14  \, u_{\nu} L^{\nu} \, \cdot\partial_\mu( f^2 L^\mu) + \frac14 f^2 L^{\nu}  \cdot dL_{\nu} \nonumber
   + \frac14 f^2 L^{\mu} u^{\nu}  \cdot \left( \partial_{\mu} L_{\nu} - \partial_{\nu} L_{\mu} \right) + u_{\nu} \partial_{\mu} \Pi^{\mu\nu}, \nonumber\\
   \label{timelike_emt_cons}
    =&\frac14 \,  u_\nu L^\nu \cdot \partial_\mu (f^2 L^\mu) + \frac{f^2}{8} dL^2  + u_{\nu}\partial_{\mu} \Pi^{\mu\nu} \, .
\end{align}
In passing to the last line we have used the structure equation 
\begin{equation}
\label{eq:equation_for_s}
\partial_{\mu} L_{\nu}-\partial_{\nu} L_{\mu} - i \left[L_{\mu},L_{\nu}\right]=0,
\end{equation}
noting that
\st
L^{\mu} \cdot 
[L_{\mu},L_\nu]
= [L^{\mu}, L_{\mu}] \cdot L_{\nu} = 0 \, .  
\stp
Adding the superfluid pressure differentials $(dp_U)_{T,\mu}$, we find after pleasing cancellations 
\begin{align}
   \left[ d\varepsilon_{U}  + (\varepsilon_U + p_U) \partial u\right]  +  (dp_U)_{T,\mu} =& 
   \tfrac{1}{2} u^{\nu}L_{\nu} \cdot  \Theta_s  +  u_{\nu}\partial_{\mu} \Pi^{\mu\nu} \, ,  \nonumber \\
   \label{tmunu_cc}
 =& -(\mu_A + \mu_A^{\diss}) \cdot  \Theta_s  +  u_{\nu}\partial_{\mu} \Pi^{\mu\nu} \, ,
 \end{align}
where we used the Josephson relation \eqref{eq:Josephsonapp} in the last step.
Combining the ingredients needed for \eqref{entropyintermediate}, from \eqref{left_cc}, \eqref{right_cc}, and \eqref{tmunu_cc}, we find 
 \begin{align}
    \partial_{\mu}(s_U u^{\mu})
=& -\frac{\mu^{\diss}_A}{T} \cdot \Theta_s 
+\frac{u_{\nu}}{T} \partial_{\mu} \Pi^{\mu \nu}  
+\frac{\mu_L}{T}\cdot \partial_{\mu}  q_L^{\mu}+ \frac{\mu_R}{T}\cdot \partial_{\mu}  q_R^{\mu} \, .
\end{align}
Integrating by parts we find finally \Eq{eq:entropyproduction} given in the text.

\section{Linearized equation for the pion field in an expanding background}
\label{pioneomapp}
In this appendix we will derive the linearized equation for the pion field for a fluid with temperature $T(x)$ and flow  $u^{\mu}(x)$, using certain hydro-kinetic approximations discussed below. 
The equation for the axial current is given by
\begin{align}
\partial_{\mu} J_{\Aa}^{\mu} - \tfrac{i}{2} [L_{\mu}, J^\mu_\Aa] 
+ \tfrac{i}{2} [L_{\mu}, J_V^{\mu} ]   =- \frac{f^2 m^2  }{4} i (U \mathcal{M}^\dagger - \mathcal{M} U^\dagger), 
\end{align}
and constitutive relations read
\begin{align}
   J_{\Aa}^{\mu} &=  \chi_A^{\rm nrm} \,  \mu_\Aa  \, u^{\mu}  + \half f^2 L^{\mu} +  q_A^\mu + \xi_A^{\mu} \, ,   \nonumber \\ 
   -\tfrac{1}{2} u^{\mu} L_{\mu} &= \mu_A +  \mu_A^{ \diss} + \xi_{\mu_A}^{\diss} \, . \nonumber
\end{align}
 Setting the isospin current $J_{V}^{\mu}$  to zero, writing 
 $U = e^{2i\varphi}$ so that $-\tfrac{1}{2} L_{\mu} \simeq \partial_{\mu} \varphi$, 
 the equations of motion to linear order in $\mu_A$ and $\varphi$  
 can be written
 \begin{gather}
    \partial_{\mu} (\chi_A^{\rm nrm} \mu_A u^{\mu})  + \partial_{\mu} (f^2 \partial^{\mu} \varphi) - f^2 m^2 \varphi +\partial_{\mu}  q^{\mu}_A  + \partial_{\mu} \xi_A^{\mu} = 0,  \\
    -u^{\mu} \partial_{\mu}  \varphi= \mu_A +\mu_A^{\diss} + \xi_{\mu_A}^{\diss} , 
 \end{gather}
 where  the dissipative strains are
 \begin{subequations}
 \begin{align}
    q_{A}^{\mu} =& -T\sigma_A \Delta^{\mu\nu} \partial_{\nu} \left(\frac{\mu_A}{T}\right) \, , \\
    \mu_A^{\diss} =& \zeta^{(2)} \left( -\partial_{\mu}(f^2 \partial^{\mu}\varphi) +  f^2 m^2 \varphi \right) \, .
 \end{align}
 \end{subequations}
 We will work  to first order in the dissipative parts yielding
 \st
 -\partial_{\mu} (\chi_A G^{\mu\nu} \partial_{\nu}\varphi) + f^2 m^2 \varphi - \partial_{\mu} q^{\mu}_A + \partial_{\mu}(\chi_A^{\rm nrm} \mu^{\diss}_A u^{\mu} ) = \xi   \, ,
 \stp
 where 
 \st
 G^{\mu\nu} \equiv -u^{\mu} u^{\nu} + v^2 \Delta^{\mu\nu}   \, ,
 \stp
 is the fluid metric introduced in \eqref{eq:fluidmetric}, and we have amalgamated the noises into a generic one
 \st
 \xi = -\partial_{\mu} (\chi^{\rm nrm}_A \xi_{\mu_A}^{\diss} u^{\mu})  + \partial_{\mu}  \xi_{A}^{\mu}  \, .
 \stp

 We  will neglect the space time derivatives of the background temperature and
 flow velocity in the dissipative terms (which are already small), but keep 
 the gradients in the ideal terms.
 Indeed, denoting the mean free path $\frac{\sigma}{\chi_A} \sim  \lambda$, the typical fluid gradient   $\partial u \sim 1/L$,
and the pion derivative $\partial \varphi \sim m \varphi$,
the different terms in the equation of  motion  are of order :
 \st
     \partial^2 \varphi \sim m^2 \varphi, \qquad 
    (\partial u) (\partial \varphi) \sim  \frac{m}{L} \varphi,  \qquad
    \frac{\sigma}{\chi_A} \partial^3\varphi \sim  \, \lambda m^3 \varphi,\
    \qquad 
    \frac{\sigma}{\chi_A} (\partial u) \partial^2 \varphi \sim  \, \frac{\lambda}{L} m^2 \varphi \, ,
    \stp
    up to an overall factor of $\chi_A$.
    In the hydro-kinetic approximation of \cite{Akamatsu:2016llw,An:2019osr}, we have
\st
(\lambda/L) m^2  \ll 
(m/L)  \sim \lambda m^3  \ll m^2 \, .
\stp
We note in passing that the neglected dissipative coefficient $\zeta^{(1)}$ gives a correction to the equation of motion of order
\st
\zeta^{(1) } (\partial^2\varphi) \, (\partial u)  \sim   (\lambda/L) m^2 \, ,
\stp
which should be dropped in our approximation scheme.

With these approximations we have
 \begin{align}
    -\partial_{\mu} q^{\mu}_A &\simeq -\sigma_A  \,  \nabla^2_\perp \; \partial_{\tau} \varphi ,  \\
    \partial_{\mu} (\chi_A^{\rm nrm} \mu_A^{\diss} u^{\mu} ) &\simeq  \chi_A^{\rm nrm} \zeta^{(2)} f^2  \, \left(\partial_{\tau}^3 \varphi -  \nabla^2_\perp \partial_{\tau} \varphi   +   m^2 \, \partial_{\tau} \varphi \right) . \ 
 \end{align}
 Here we have defined various derivatives in the rest frame
 \begin{align}
 \partial_{\tau} \varphi \equiv u^{\mu}\partial_{\mu}\varphi,   \qquad \nabla_\perp^{\mu}\equiv  \Delta^{\mu\nu}  \partial_{\nu} ,
 \qquad \nabla^2_\perp \varphi \equiv \Delta^{\mu\nu} \partial_{\mu} \partial_{\nu} \varphi,   \qquad \partial^{\mu} = -u^{\mu}  \partial_{\tau} + \nabla_\perp^{\mu},
 \end{align}
 which all commute when approximating  the dissipative currents.
 Next we use the lowest order equations of motion 
\begin{equation}
   \label{lowesteom}
\partial_\tau^2 \varphi = \frac{f^2}{\chi_A} \nabla^2_\perp \varphi  - \frac{f^2m^2 }{\chi_A}\varphi  , 
\end{equation}
 to rewrite the triple  time derivative 
 \begin{align}
    \partial_{\mu} (\chi_A^{\rm nrm} \mu_A^{\diss} u^{\mu} ) &\simeq  
    -(\chi_A^{\rm nrm} v)^2 \zeta^{(2)}  \,    \nabla^2_\perp \partial_{\tau} \varphi
             +    (\chi_A^{\rm nrm})^2 \zeta^{(2)}\, v^2 m^2 \, \partial_{\tau} \varphi  \, .
 \end{align}
With these steps,  the wave equation can be written
 \st
 -\partial_{\mu} (\chi_A G^{\mu\nu} \partial_{\nu}\varphi) + f^2 m^2 \varphi  - \lambda_A \nabla^2_\perp \partial_\tau \varphi + \lambda_{m} m^2  \, \partial_\tau \varphi = \xi  \, ,
 \stp
 where
\begin{align}
   \label{coeffsdef}
\lambda_A &\equiv (\chi_A^{\rm nrm} v)^2 \zeta^{(2)} +\sigma_A, \\
\lambda_{m} &\equiv (\chi_A^{\rm nrm}v)^2 \zeta^{(2)}  .
\end{align}
which is the form used in the text \eqref{wave_eq} and the Appendix~\ref{app:kin}, see \eqref{eq:pioneomapp}.

Finally, let's check the form of the noise term,  
using the same approximation scheme. 
In Fourier space, with four vector $q_{\mu}$,
the correlation function of the noise takes the form
\begin{align}
   \label{noisecorrelatorinq}
   \llangle \xi(q) \xi(-q) \rrangle \simeq 2 T\zeta^{(2)} (\chi_A^{\rm nrm})^2 (u^{\mu} q_{\mu})^2 + 2T \sigma_A \Delta^{\mu\nu} q_{\mu} q_{\nu} \, ,
\end{align}
where we have used the noise correlators given in \eqref{variancesfundamental}.
Next, we should recall that we are close to being on-shell where 
$(u^{\mu} q_{\mu})^2 =  v^2(\Delta^{\mu\nu} q_{\mu} q_{\nu} + m^2)$. (This on-shell relation is \eqref{lowesteom} written in Fourier space.)   Inserting the on-shell relation  into \eqref{noisecorrelatorinq}, and taking the Fourier transform, shows noise correlator can be written
\begin{align}
\label{eq:noisecorrelatorapp}
   \llangle \xi(x) \xi(y) \rrangle \simeq 2 T \left( -\lambda_A \nabla_\perp^2 + \lambda_{m} m^2 \right) \delta(x -y) \, .
\end{align}
This completes the derivation of \eqref{wave_eq} given in the text.

\section{Hydro-kinetic transport equation for soft pions at zero isospin density }\label{app:kin}

 The equations of superfluid hydro describe how soft Goldstone modes (i.e. pions)  interact 
 with the stress tensor of the normal fluid.  Since the wavelength of these
 modes is short compared to the wavelengths associated the energy momentum tensor, the evolution of the pion modes is described by a Boltzmann equation. 
 The stochastic superfluid hydrodynamic theory can be used to determine the form of this Boltzmann equation (which looks like a relaxation time equation), in much the same way that the kinetic equations for sound modes can be determined from stochastic hydrodynamics~\cite{Akamatsu:2016llw,An:2019osr}. Our 
 goal here is to derive  the results of \Sect{sec:pionkinetics}. Good derivations of the Boltzmann equation from a stochastic wave equation can be found in several places~\cite{Arnold:1998cy,kamenev2011field,PhysRevD.78.085027}. Here we will follow \cite{kamenev2011field}. 

 \subsection{Derivation of the Boltzmann equation}

 We will derive the transport equation in the absence of a net isospin charge. 
In this case each component of $\varphi_{a}$  is independent, and the distribution function $f_{ab}$ is diagonal. We will therefore
derive the transport equation for a one component scalar field.  The wave equation for the pion fluctuations takes the form (see Appendix~\ref{pioneomapp} for definitions)
\st
\label{eq:pioneomapp}
-\partial_{\mu} (\chih \, G^{\mu\nu} \partial_{\nu} \varphi) + f^2 m^2 \varphi    -\lambda_{A}  \nabla_\perp^2 \partial_{\tau} \varphi  + \lambda_m  m^2 \partial_\tau \varphi = \xi \, ,
\stp
Here the parameters,  $G^{\mu\nu}$, $u^{\mu}$, $f^2$, $m^2$, $\lambda_A$ and $\lambda_{m}$
depend slowly on space and time,   
and the variance of the noise is
given in \eqref{eq:noisecorrelatorapp}.
The gradients drive the pion distribution weakly out of equilibrium, while the dissipation and noise tries to re-establish local equilibrium.

Our goal is to derive  the kinetic equation associated with this stochastic
wave equation by making the appropriate quasi-particle  approximations.
The first two terms come from the ideal equations of
motion, while the last two terms are viscous corrections.   Space-time gradients
to the ideal equations of motion  are of the same order as viscous corrections and
will be included in developing the transport equations.  However,  space-time
gradients to the viscous parts of the equations of motion are smaller and will be ignored (see Appendix~\ref{pioneomapp}).  

To streamline the discussion we introduce the following linear operator  with 
retarded boundary conditions:
\st
\mathcal L_{xy} \equiv \left[ -\partial_{\mu} ( \chih \, G^{\mu\nu} \partial_{\nu} ) +
   f^2 m^2 \varphi_a  -   \lambda_A \nabla_\perp^2 \partial_\tau 
 + \lambda_{m}  m^2  \,\partial_\tau 
  \right] \delta(x-y)   \, . 
\stp
Here it is understood that the parameters (such as $G^{\mu\nu}(x)$) are functions of the space-time coordinates $x^{\mu} = (x^0, {\bm x})$. 
Below we will employ a hyper-condensed notion where repeated coordinates are
integrated over, e.g.
\st
G_R(x,y) \varphi(y) \equiv \int d^4y \, G_R(x,y)\, \varphi(y) \, ,
\stp
The retarded Green function satisfies
\st
\mathcal L_{xz} \, G_R(z,y) = \delta(x-y) \, ,
\stp 
while the advanced Green function satisfies 
\st
\mathcal L_{yz}\, G_A(x,z) = \delta(x -y) \, .
\stp
The equations of motion are thus 
\st
\mathcal L_{xx'} \varphi(x') =   \xi(x),
\stp
and the two point functions satisfy
\st
\mathcal L_{xx'}  \,  \mathcal L_{yy'} \llangle \varphi(x') \varphi(y') \rrangle =  \llangle \xi(x) \xi(y) \rrangle.
\stp

The distribution function $N(x,y)$ is defined (see below for motivation) from the symmetrized two point functions of fields via an integral equation~\cite{kamenev2011field}
\st
\label{pinchdef}
\llangle \varphi(x) \varphi(y)  \rrangle = -i  \left(G_{R}(x,z) N(z,y)  - N(x,z) G_{A}(z,y) \right) \, ,
\stp
With this definition, $N(x,y)$ evolves as
\st
\label{eomforN}
i \left(\mathcal L_{xz} N(z,y) - \mathcal L_{yz} N(x,z) \right) =  \llangle \xi(x) \xi(y) \rrangle.
\stp

Now we will make a Wigner transform, defining average $\bx = (x + y)/2$ and 
difference $s = x -y$ coordinates.
The Wigner transform takes the form 
\begin{multline}
   \int d^4s \, e^{- ip \cdot s} \,  A(x,z) \, B(z,y)  =  
A(\bx,p) B(\bx,p)  \\
+ \frac{i}{2}  \left( 
   \frac{\partial A(\bx,p)}{\partial \bx^{\mu}}  \frac{\partial B(\bx,p)}{\partial p_{\mu}} -  
\frac{\partial A(\bx,p)}{\partial p_{\mu}}  \frac{\partial B(\bx,p)}{\partial \bx^{\mu}}  \right)   + \ldots 
\end{multline}
The Wigner transform of the  differential operator takes the form
\st
\int d^4s \, e^{-i p \cdot s} \mathcal L_{xy}   = 2 \chih(\bx) \,  \left( \F(\bx,p) - i  \,  E(\bx ,p)\,\Gamma_p(\bx,p)/2 \right)  \,   .
\stp

Here the ``Hamiltonian'' is 
\st
\F(x,p) = \frac{1}{2} G^{\mu\nu}(x) \, p_{\mu} p_{\nu} + \frac{1}{2} v^2(x) \, m^2(x)   \, ,
\stp
where the rest frame energy is $E(\overline{x},p) \equiv -u^{\mu} p_{\mu}$, and the quasi-particle damping rate is $\Gamma_p(x,p) = D_A \, \Delta^{\mu\nu} p_{\mu} p_{\nu} + D_m \, m^2$.
The retarded Green function is the inverse of $\mathcal L_{xy}$
\st
\label{RetardedG}
G_{R}(\bx,p)  \simeq  \frac{1}{2 \chih } \, \frac{1}{\F - i \, E \, (\Gamma_p/2) } = \frac{1}{\chih} \, \frac{1}{\left(-E^2 + \omega_p^2 - i E \, \Gamma_p\right)} \, ,
\stp
where  $\omega_p^2 = v^2 (\Delta^{\mu\nu}p_{\mu} p_{\nu} + m^2)$ is the quasi-particle energy in the rest frame.

We now will give present the motivation for the definition  of
the distribution function based on \eqref{pinchdef}.   
The first motivation comes by considering equilibrium. In this case we may
use full Fourier transforms and use translational invariance $ \llangle \varphi(x) \varphi(y)  \rrangle \equiv G_{\rm sym}(x-y) $,  
where the symmetrized distribution correlation function as a function of momentum is
\st
G_{\rm sym}(p)  =  -i\left( G_R(p) - G_A(p) \right) N(p) \, .
\stp
In order to satisfy the  fluctuation dissipation theorem,  we must have
\st
N(p)  \rightarrow  n(E) + \frac{1}{2} \simeq \frac{T}{E} \, ,
\stp
in equilibrium.  

The next motivation
comes from taking averages  of the fields. Consider
the average 
\st
\label{pinched}
\llangle \partial_{\mu} \varphi(x) \partial_{\nu} \varphi(x) \rrangle 
\simeq  \frac{1}{2\chih } \int \frac{d^4p}{(2\pi)^4}  p_{\mu} p_{\nu} N(x,p) \left[ \frac{-i}{\F - i \, E \,(\Gamma_p/2)}   -  \frac{-i}{\F + i\,  E \,(\Gamma_p/2) } \right] \, .
\stp
The presence of the difference between the retarded and advanced propagators 
means that the  integration over $p_{0}$ is ``pinched'' whenever $\mathcal H$ 
approaches zero, i.e. whenever the particle goes on-shell. Using the pinch approximation, we find
\st
 \llangle \partial_{\mu} \varphi(x) \partial_{\nu} \varphi(x) \rrangle 
 \simeq  \frac{1}{2\chi_A}\int \frac{d^4p}{(2\pi)^4} \, 2\pi \delta(\F) \, {\rm sign}(E)\, p_{\mu} p_{\nu}\, N(x,p)\, .
\stp
The $\delta$-function is satisfied at two roots $p_0=-h_{\pm}(x,p_i)$, 
and we write
\st
2\pi \delta(\mathcal H) =   \frac{2\pi }{|\partial \mathcal H/\partial p_0|} \delta(p_0 + h_+(x,\p))  +  \frac{2\pi}{|\partial \mathcal H/\partial p_0|} \delta(p_0 + h_-(x,\p)) \, ,
\stp
where $\partial \mathcal H/\partial p_0 = G^{0\nu} p_{\nu}$, and 
\st
\label{hpmonshell}
h_{\pm} (x,p_i) =\frac{G^{0i}p_i} {G^{00}} \pm \frac{1}{\sqrt{-G^{00}}} 
\sqrt{
\left( G^{ij} + \frac{G^{0i} G^{0j}  }{-G^{00}}  \right) p_i p_j + v^2 m^2}  .
\stp
Note that $h_{-}(x,-\p) =  -h_+(x,\p)$.  For future reference we also note that
$ E = - p^{\mu} u_{\mu} = \pm \omega_p$.

The integral  in \eqref{pinched} breaks up 
into a positive piece and negative piece. After changing variables $\p \rightarrow -\p$ in the negative piece,
the integral takes  the form
\st
 \llangle \partial_{\mu} \varphi(x) \partial_{\nu} \varphi(x) \rrangle  
 \simeq  \frac{1}{2 \chih} \, \int \frac{d^3p_i}{(2\pi)^3 \,  (\partial \mathcal H/\partial p_0) } \,   p_{\mu}  p_{\nu} \left[ N(x,   p) -  N(x,- p) \right],
\stp
where now  the momentum is evaluated on the positive mass shell
\st
p_{\mu} = (-h_+(x,p_i), p_i), \quad  \mbox{with} \quad    \frac{\partial \mathcal H}{\partial p_0 }  > 0\, .
\stp
For real fields $N(x,p) =  - N(x,-p)$, so 
\st
\label{onshellaverage1}
\llangle \partial_{\mu} \varphi(x) \partial_{\nu} \varphi(x) \rrangle 
\simeq \frac{1}{\chih}  \int \frac{d^3p_i}{(2\pi)^3  (\partial \mathcal H/\partial p_0)}  \,   p_{\mu}   p_{\nu} \, N(x,  p).
\stp

With these preliminaries we can determine the equation of motion $N(x, p)$ on mass shell. The Wigner transform of \eqref{eomforN} yields an
equation of motion for $N(x,p)$  of the form
\st
 \frac{\partial (\chih \mathcal H)}{\partial p_{\mu} }    \frac{\partial N(x,p)}{\partial x^{\mu}} - \frac{\partial (\chih \mathcal H)}{\partial x^{\mu} } 
\frac{\partial N(x,p)}{\partial p_{\mu}} 
= - \, \chih\,  \Gamma_p \left[ E N(x,p)  - T \right].
\stp
As a first step towards
putting the distribution on shell (with $p^0$ or $E$ positive), we define $f(x, p_i, \mathcal H)$,
which is parameterized by $\mathcal H$ instead of $p_0$ 
\st
N(x,p_0,p_i) \equiv  f_\pi(x,p_i,\chih \F)  \, .
\stp
The equation of motion for $f_\pi(x,p_i, \chih \mathcal H)$,  simply loses the $\partial /\partial p_0$ term since the Poisson bracket of $\chih\F$ with itself is zero:
\st
\frac{\partial (\chih \F) }{\partial p_{\mu} } 
\frac{\partial f_\pi}{\partial x^{\mu}} - \frac{\partial (\chih\F)}{\partial x^{i} } 
\frac{\partial f_\pi}{\partial p_{i} } 
= - \chih \Gamma_p \left[ E f_\pi  - T \right] \, ,
\stp
In evaluating equal time expressions of fields as in \eqref{onshellaverage1}, we only need  
the distribution evaluated on-shell where $\F=0$ and $E = \omega_p$ , yielding the equation of motion given in the text \eqref{transporteqnfinal}. 
We also note that the velocity  and force of the soft pions is given by
\begin{align}
\frac{ \partial \mathcal H/\partial p_i } {
\partial \mathcal H/\partial p_0 } &= \frac{\partial h_+(x,p_i) }{\partial p_i} , \\
-\frac{ \partial \mathcal H/\partial x^i } {
\partial \mathcal H/\partial p_0 } &= -\frac{\partial h_+(x,p_i) }{\partial x^i}  ,
\end{align}
leading to an alternate form of the Boltzmann equation \eqref{transporteqnfinalb}.

\subsection{Derivation of the Boltzmann stress tensor}
\label{BoltzStress}

To complete the Boltzmann picture we need to evaluate the stress tensor. We have 
already discussed how to evaluate stochastic averages such as $\llangle \partial_{\mu} \varphi \partial_{\nu} \varphi \rrangle$, with the result
\st
\label{onshellaverage}
\chi_A \llangle \partial_{\mu} \varphi(x) \partial_{\nu} \varphi(x) \rrangle 
\simeq   \int \frac{d^3p_i}{(2\pi)^3  (\partial \mathcal H/\partial p_0)}  \,   p_{\mu}   p_{\nu} \, f_{\pi}(x,p_i) \, .
\stp

Expanding the  superfluid stress tensor given in \eq{Tmunudef} to quadratic order in $\varphi$ and  $\mu_A$  with $\mu_V=0$,  and then averaging over the stochastic fluctuations of the pion field yields  the coarse grained stress tensor:
\st
\llangle T^{\mu\nu}(x) \rrangle  = e(T) u^{\mu} u^{\nu} + p(T) \Delta^{\mu\nu}  +  T^{\mu\nu}_\pi ,
\stp
where the pion contribution  is\footnote{Recall that in this appendix $\varphi$ denotes one isospin component of the  pion field.
   We have multiplied \eq{Tmunustart} by $d_A=3$ to account for
   the three pion states.
}
\st
\label{Tmunustart}
 T^{\mu\nu}_\pi = d_A \llangle (e_{\varphi} + f \mu^2 ) u^{\mu} u^{\nu} + p_{\varphi} \Delta^{\mu\nu} 
+ f^2 \Delta^{\mu\alpha} \Delta^{\nu\beta} \partial_{\alpha} \varphi \partial_{\beta} \varphi \rrangle  \, .
\stp
Here the axial chemical potential is $\mu = -u^{\mu} \partial_\mu \varphi$,  the pressure to quadratic order is 
\st
 p_\varphi 
\equiv  -\chih \left( \frac{1}{2} G^{\mu\nu} \partial_{\mu}\varphi  \partial_{\nu} \varphi + \frac{1}{2} v^2 m^2 \varphi^2  \right) \, ,
\stp 
and the energy density is 
\st
 e_\varphi  \equiv  -p_{\varphi} + T \frac{\partial p_{\varphi}}{\partial T} + \mu \frac{\partial p_{\varphi}}{\partial \mu}  \, .
\stp
The pressure $p_\varphi$ is a function of  $T,\mu, (\partial\varphi)^2$ and $\varphi^2$, echoing the discussion surrounding \eqref{epsilonU}.

Now let us evaluate $T^{\mu\nu}_\pi$ in a kinetic approximation.
The $ \llangle \Delta^{\mu\alpha} \Delta^{\nu\beta} \partial_\alpha \varphi \partial_\beta \varphi \rrangle $ 
term leads to  the second term in \eqref{Tmunuboltz}. 
The pressure $p_{\varphi}$ is closely related to the Hamiltonian $\mathcal H$, and we find
\begin{subequations}
\begin{align}
  \llangle p_\varphi (x)\rrangle &= \int \frac{d^3 p_i }{(2\pi)^3\,  (\partial \mathcal H/\partial p_0) } \, \mathcal H(x,p) \, f_{\pi}(x,p) ,  \\
   &= 0 , 
\end{align}
\end{subequations}
since $\mathcal H(x,p) = 0 $ on-shell. 
Finally, careful algebra together with the constraint $\mathcal H(x,p) = 0$ yields
\st
\llangle e_{\varphi} + f\mu^2 \rrangle =  \int \frac{d^3 p_i }{(2\pi)^3\,  (\partial \mathcal H/\partial p_0) } \, \omega_p \frac{\partial (\beta \omega_p)}{\partial \beta} \, f_{\pi}(x,p)  \, .
\stp
Putting together the ingredients leads to \eqref{Tmunuboltz}.

\end{appendix}

\bibliography{superpaper1v2}

\end{document}